\documentclass[aps,showpacs,nofootinbib,superscriptaddress,floatfix]{revtex4} % for mac
\usepackage{amsmath}
\usepackage{graphicx,epsfig,wrapfig}
\usepackage{amsthm}
\usepackage{amsfonts}
\usepackage{amssymb}
\usepackage{natbib}
\usepackage{eufrak}
\usepackage{dsfont}
\usepackage{mathrsfs}
\usepackage{slashed}
\usepackage{epstopdf}
\usepackage{bm}
\usepackage{color}
\usepackage[normalem]{ulem} 
\renewcommand\sout{\bgroup \color[rgb]{0.55,0.00,0.99} \ULdepth=-.5ex \ULset}

\begin{document}
\newcommand{\red}[1]{{\color{red} #1}}
\newcommand{\blue}[1]{{\color{blue} #1}}
\newcommand{\mb}[1]{\mathbf{#1}}
\newcommand{\mc}[1]{\mathcal{#1}}
\newcommand{\non}{\nonumber}
\newcommand{\pep}[1]{\mathbf{#1}_{\perp}}
\newcommand{\pepgr}[1]{\bm{#1}_{\perp}}
\newcommand{\gdir}[1]{\gamma^{#1}}
\newcommand{\xk}{(x,\mathbf{k}_{\perp})}
\newcommand{\xkq}{(x,\mathbf{k}_{\perp}^{2})}
\newcommand{\pe}{(\mb{p}_{e})}
\newcommand{\xks}{(x,\pep{k};S)}
\newcommand{\pv}{\mathrm{P.V.}}
\newcommand{\xkl}{(x,\pep{k};\Lambda)}
\newcommand{\vecperp}[2] {\mathbf{#1}_{\perp}^{#2}}
\newcommand{\myvec}[1] {\mathbf{#1}}
\newcommand{\piplus}{\pi^{+}}
\newcommand{\piminus}{\pi^{-}}
\newcommand{\myprime}[2]{#1^{\prime #2}}
\newcommand{\mybar}[2]{\bar{#1}^{#2}}
\newcommand{\myref}[1]{\( (\ref{#1}) \)}
\def\bra#1{\mathinner{\langle{#1}|}}
\def\ket#1{\mathinner{|{#1}\rangle}}
\def\braket#1{\mathinner{\langle{#1}\rangle}}
\newcommand{\gev}{~\mbox{GeV}^2}

\newcommand{\ux}{\boldsymbol{x}}
\newcommand{\uy}{\boldsymbol{y}}
\newcommand{\uz}{\boldsymbol{z}}
\newcommand{\ur}{\boldsymbol{r}}
\newcommand{\un}{\boldsymbol{n}}
\newcommand{\up}{\boldsymbol{p}}
\newcommand{\uP}{\boldsymbol{P}}
\newcommand{\uq}{\boldsymbol{q}}
\newcommand{\uQ}{\boldsymbol{Q}}
\newcommand{\uS}{\boldsymbol{S}}
\newcommand{\us}{\boldsymbol{s}}
\newcommand{\uK}{\boldsymbol{K}}
\newcommand{\uzero}{\boldsymbol{0}}
\def\nn{\nonumber}

\def\dirac#1{\slash \mkern-10mu #1}
\def\ket#1{\hbox{$\vert #1\rangle$}}   % definition of ket
\def\bra#1{\hbox{$\langle #1\vert$}}   % definition of bra

\newcommand{\bea}{\begin{eqnarray}}
\newcommand{\eea}{\end{eqnarray}}
\newcommand{\be}{\begin{equation}}
\newcommand{\ee}{\end{equation}}

\newcommand{\ud}{\mathrm{d}}
\newcommand{\uL}{\mathcal{L}}
\newcommand{\uM}{\mathcal{M}}
\newcommand{\uPcal}{\mathcal{P}}
\newcommand{\uZ}{\mathcal{Z}}
\newcommand{\uQcal}{\mathcal{Q}}
\newcommand{\uN}{\mathcal{N}}
\newcommand{\uTr}{\mathrm{Tr}}
\newcommand{\uSp}{\mathrm{Sp}}
\newcommand{\uslash}{/\!\!\!}
\newcommand{\dslash}{\partial\!\!\!/}
\newcommand{\uk}{\boldsymbol{k}}
\newcommand{\uD}{\mathcal{D}}
\newcommand{\uV}{\boldsymbol{V}}
\newcommand{\uVcal}{\mathcal{V}}
\newcommand{\sgn}{\text{sgn}}

% new macro for bras, kets and brakets
\def\bra#1{\mathinner{\langle{#1}|}}
\def\ket#1{\mathinner{|{#1}\rangle}}
\def\braket#1{\mathinner{\langle{#1}\rangle}}

\title{Collinear parton distributions and the structure of the nucleon sea  in a light-front meson-cloud model}

\author{Stefan Kofler}
\email{stefan.kofler@uni-graz.at}

\affiliation{Institut f\"ur Physik, Universit\"at Graz, 8010 Graz, Austria}
\author{Barbara Pasquini}
\email{barbara.pasquini@unipv.it}

\affiliation{Dipartimento di Fisica, Universit\`a degli Studi di Pavia, I-27100 Pavia, Italy}
\affiliation{Istituto Nazionale di Fisica Nucleare, Sezione di
  Pavia,  I-27100 Pavia, Italy}

\date{\today}


\allowdisplaybreaks[2]

\begin{abstract}
The unpolarized, helicity and transversity parton distribution functions of the nucleon are studied within a convolution model where the bare nucleon is dressed by its virtual meson cloud. 
Using light-front time-ordered perturbation theory, the Fock states of the physical nucleon are expanded in a series involving a bare nucleon and two-particle, meson-baryon, states.
The bare baryons and mesons are described with light-front wave functions (LFWFs) for  the corresponding valence-parton components. Using a representation in terms of overlap of LFWFs, the role of   the non-perturbative antiquark degrees of freedom and the valence quark contribution at the input scale of the model is discussed for the leading-twist collinear parton distributions. 
After introducing perturbative QCD effects through evolution to experimental scales, the results  are compared with available  data and phenomenological extractions.
Predictions for the nucleon tensor charge are also presented, finding a very good agreement with recent phenomenological extractions.
\end{abstract}

\maketitle
\section{Introduction}%\renewcommand{\arraystretch}{1.5}
A successful approach in high energy physics to describe the partonic structure of the nucleon 
is based on light-front quantization, where hadrons are described in terms of light-front wave functions (LFWFs)~\cite{Brodsky:1997de,Bakker:2013cea}. Representation of parton distribution functions  in terms of overlap integrals of LFWFs has proven to be a powerful framework to unveil the underlying physical picture and provide the support for theoretical modeling. 
At fixed light-front time, the nucleon state can be decomposed
 in terms of various quark ($q$), antiquark ($\bar q$) and gluon $(g)$ Fock components, i.e.
\bea
|N\rangle=\psi_{(3q)}|qqq\rangle+\psi_{(3q+q\bar q)}|qqqq\bar q\rangle+\psi_{(3q+1g)}|qqqg\rangle+\dots\, , 
\eea
where the LFWFs $\psi_{(\dots)}$ represent the probability amplitudes  to find the different parton configurations in the nucleon.
A general model-independent classification of the LFWFs for the $3q$ and $3q+1g$ components has been worked out in Refs.~\cite{Ji:2002xn,Ji:2003yj}. 
However, to probe the parton content of the nucleon suitable models have to be invented to give explicit expressions for the LFWFs.
Most of the applications have focused on the minimum parton content, providing a description of the  valence-quark contribution to the leading-twist parton distribution functions entering in various deep inelastic scattering processes~\cite{Lorce:2011dv,Burkardt:2015qoa}.
A  step forward to include also the Fock state with one additional gluon has been performed in Ref.~\cite{Braun:2011aw}, allowing one to extend the discussion to higher-twist parton distribution functions.
Recently, works have been done to describe also the non-perturbative structure of the nucleon sea encoded in the $3q+q\bar q$ component of the LFWF, by integrating meson-cloud effects into the valence-quark contribution.
Along the lines originally proposed in Refs.~\cite{Drell:1969wd,Sullivan:1971kd}, a meson-baryon Fock-state expansion is used to construct the state $|\tilde N\rangle$ of a dressed physical nucleon. In the one-meson approximation the state $|\tilde N\rangle$ is pictured as being part of the time a bare nucleon, $|N\rangle$, and part of the time a baryon-meson system, $|B, M\rangle$.
Using light-front quantization to resolve the structure of the nucleon core $|N\rangle$ and of the  bare meson and baryon in the $|B,M\rangle$ state in terms of the constituent partons,
one can build up the corresponding $3q$ and $3q+q\bar q$ components of the LFWFs.
Explicit expressions for the LFWFs within a light-front meson-cloud model have been constructed in Refs.~\cite{Pasquini:2006dv,Pasquini:2007iz}, with applications to the description of the leading-twist unpolarized generalized parton distributions (GPDs)~\cite{Pasquini:2004gc,Pasquini:2006dv}, electroweak form factors of the nucleon~\cite{Pasquini:2007iz}, and nucleon to pion transition distribution amplitudes~\cite{Pasquini:2009ki}.
More recently, a detailed study  has been presented in Refs.~\cite{Traini:2013zqa,Traini:2011tc} for the unpolarized parton distribution function (PDF) within a light-front meson-cloud model including perturbative effects up to next-to-next-to-leading-order (NNLO) accuracy.
These works enter in the class of an extensive literature within the meson-cloud model (see Refs.~\cite{Speth:1996pz,Londergan:1998ai,Chang:2014jba,Kumano:1997cy}, and references therein), by introducing the novel approach of a fully relativistic light-front formalism for the description of both the kinematics of the $N\rightarrow BM$ fluctuations and the quarks dynamics encoded in the LFWFs of the core baryon and meson states.
 As a matter of fact, the meson-cloud model has received a great deal of attention, since it was realized that it can give an explanation of the flavor-symmetry violation in the sea-quark distributions of the nucleon~\cite{Thomas:1983fh} by accounting for the excess of $\bar d$ antiquarks over $\bar u$ antiquarks as observed through the violation of the
Gottfried sum rule~\cite{Amaudruz:1991at,Arneodo:1994sh}. 
Although the nucleon's nonperturbative antiquark sea cannot be ascribed entirely to its virtual meson-cloud~\cite{Koepf:1995yh}, the role of mesons in deep inelastic scattering (DIS) is of primary importance~\cite{Vogt:2000sk}.

In this work, we review the application of the meson-cloud model within light-front quantization to the leading-twist PDFs.
This approach has been recently discussed in the case of the unpolarized PDF~\cite{Traini:2013zqa,Traini:2011tc} and will be extended here to consider also the longitudinally and transversely polarized PDFs. In particular, the transversity distribution has never been discussed 
so far in the context of meson-cloud models, and we will present here for the first time the 
convolution formalism to account for the sea quark contribution to transversity within a meson-cloud picture of the nucleon. The calculation is performed using LFWFs for the bare nucleon and bare mesons, which have proven to give a faithful description of  the core structure of the hadrons 
as probed in various observables. 
Another important ingredient that will be discussed is the matching scale of our hadronic model consistent with QCD evolution. Once  the input scale of the model is identified, we will be able to apply evolution equations  to evolve our results at the relevant scale of experiments. 

The paper is organized as follows: In Sec.~\ref{sect:2} the relevant formulas for the LFWF of the dressed nucleon 
in the  meson-cloud model are collected. The convolution formalism for the calculation of the  three leading-twist collinear PDFs within the meson-cloud model is discussed in Sec.~\ref{sect:3}, while the ingredients for the explicit  calculation of the PDFs in terms of overlap of LFWFs are presented in Sec.~\ref{sect:4}. In Sec.~\ref{sect:5}, after fixing the input scale of the model, we present our results for the valence-quark and sea-quark contributions at the hadronic scale of the model as well as after leading-order (LO) evolution to the experimental scales. We then compare our findings with available experimental measurements and phenomenological extractions. Finally, we discuss our results for  the tensor charge of the nucleon in comparison with various theoretical calculations and recent phenomenological extractions.
In Sec.~\ref{sect:6} we summarize our conclusions.
Technical details necessary to derive the convolution formulas for the PDFs in the meson-cloud model are given in the App.~\ref{appendix:a}.

  \section{The meson-cloud model of the nucleon}
\label{sect:2}The basic assumption of the meson-cloud model is that the physical nucleon state \(\ket{\tilde{N}} \)
can be expanded (in the infinite momentum frame (IMF) and in the one-meson
approximation) in a series involving  a bare nucleon \( \ket{N} \)  and two-particle,
meson-baryon states \( \ket{B,M} \).
The wave function of the physical nucleon is then expanded in terms of the bare nucleon and meson-baryon Fock states, i.e.
\begin{eqnarray}
&&|\tilde p_N,\lambda;\tilde N\rangle
=\sqrt{Z}|\tilde p_N,\lambda; N\rangle+
\sum_{B,M}
\int \frac{{\rm d}y{\rm d}^2{\bf k}_{\perp}}{2(2\pi)^3}\,
\frac{1}{\sqrt{y(1-y)}}
\sum_{\lambda',\lambda''}
\phi_{\lambda'\lambda''}^{\lambda \,(N,BM)}(y,{\bf k}_\perp)
|\tilde p_B,\lambda';B\rangle\, 
|\tilde p_M, \lambda'';M\rangle,
\label{eq_physnuc_fock_state}
\end{eqnarray}

\noindent where the light-front momenta of the baryon, $\tilde p_B=(p_B^{+},\mathbf p_{B\perp})$  and the meson, $\tilde p_M=(p_{M}^{+},\mathbf p_{M\perp} )$, can be written in terms of the intrinsic (nucleon rest-frame) variables as\footnote{Light-front coordinates of a generic four-vector
$a=(a^+,a^-,\mathbf a_\perp)$  are defined by $a^+= \frac{1}{\sqrt{2}}\left( a^0+a^3 \right)$, $ a^-=\frac{1}{\sqrt{2}}\left( a^0-a^3 \right) $ and $\mathbf{a}_\perp=(a_1,a_2)$,  in terms of the standard Minkowski four-vector-components.}
\be
\begin{array}{ll}
p^+_B= y p_N^+,\qquad &p^+_M=(1-y)p_N^+,\\
{\bf p}_{B\perp}={\bf k}_\perp+y\,{\bf p}_{N\perp},
\quad &{\bf p}_{M\perp}=-{\bf k}_\perp+(1-y)\,{\bf p}_{N\perp}.
\end{array}
\label{eq:9}
\ee
In Eq.~\eqref{eq_physnuc_fock_state} we introduced the function $\phi_{\lambda'\lambda''}^{\lambda\,(N,BM)}(y,{\bf k}_\perp)$ to define the probability amplitude for a nucleon with helicity $\lambda$ to fluctuate into 
a virtual $BM$ system with the baryon  having helicity $\lambda'$, longitudinal momentum fraction $y$ and transverse momentum ${\bf k}_\perp$, 
and the meson having helicity $\lambda''$, longitudinal momentum fraction $1-y$ and 
transverse momentum $-{\bf k}_\perp$, respectively.
As explained in Refs.~\cite{Pasquini:2006dv,Pasquini:2007iz}, this function can be calculated using time-ordered perturbation theory (TOPT) in the 
infinite-momentum frame, which is equivalent to light-front time ordered perturbation theory.
The final result reads 
\begin{eqnarray}
\phi_{\lambda'\lambda''}^{\lambda\,(N,BM)}(y,{\mathbf k}_\perp)
=\frac{1}{\sqrt{y(1-y)}}\,
\frac{V^\lambda_{\lambda'\lambda''}(N,BM)}
{M^2_N-M^2_{BM}(y,{\mathbf k}_\perp)},\,
\label{eq_def_splitting_function}
\end{eqnarray}
where $M_{BM}$ is the squared invariant mass of the baryon-meson fluctuation
\begin{equation}
M^2_{BM}(y,{\mathbf k}_\perp)\equiv\frac{M_B^2+{\mathbf k}^2_\perp}{y}
+\frac{M_M^2+{\mathbf k}^2_\perp}{1-y}.
\label{eq_def_invariant_mass}
\end{equation}
In Eq.~\eqref{eq_def_splitting_function}, $V^\lambda_{\lambda'\lambda''}(N,BM)$ is the vertex function describing the transition of the nucleon into a baryon-meson state, which has been explicitly calculated for various transitions and helicity combinations, e.g.,  in Refs.~\cite{Pasquini:2006dv,Pasquini:2007iz}.
\\
\noindent The hadron states are normalized as
\begin{equation}
\label{eq_hadron_normalization}
 \braket{p^{\prime +},\mathbf{p}_{\perp}^{\prime}, \lambda^{\prime}; H | p^{ +},\mathbf{p}_{\perp}, \lambda; H} = 2(2 \pi)^3  p^+ \delta \left(p^{\prime +} - p^+ \right) \delta^{(2)} \left(\mathbf{p}_{\perp}^\prime - \mathbf{p}_\perp \right).
\end{equation}
\\
\noindent By imposing the normalization on the hadron state of Eq.~\eqref{eq_physnuc_fock_state}, we obtain the following condition on the normalization factor $Z$
\be
1=Z+\sum_{B,M}P_{N/BM},
\label{eq_Z_factor}
\ee
with\footnote{For better legibility we denote the helicities of spin  \( 1/2 \) particles with \( \pm \) instead of $\pm\frac{1}{2}$.}
\be
P_{N/BM}=
\int \frac{{\rm d}y \, {\rm d}^2{\bf k}_{\perp}}{2(2 \pi)^3}\,
\sum_{\lambda',\lambda''}
\left|\phi^{+(N/BM)}_{\lambda'\lambda''}(y,\mathbf{k}_\perp)\,\right|^2.
\label{eq_probabilities}
\ee
Here
$P_{N/BM}$ is the probability that a nucleon fluctuates into a baryon-meson state, and, accordingly, $Z$ gives  the probability to 
find the bare
nucleon in the physical nucleon. 
 \section{Convolution model for the parton distribution functions}
  \label{sect:3}
According to the Sullivan description~\cite{Sullivan:1971kd}, in a DIS process  there are no interactions among the particles in a multiparticle
Fock state during the interaction with the hard photon.
Therefore, the virtual photon can hit either the bare proton\footnote{We can restrict our discussion about the PDFs to the proton, since the PDFs of the neutron can be related to those of the proton by isospin invariance.}
$p$  or one of the constituents of the higher Fock states. 
As a consequence,  a generic quark parton distribution functions $j(x)$ can be obtained by the sum of two contributions
\begin{equation}
j^{q/p} (x) = Z j^{q_V/p}_{bare}(x) + \delta j^{q}(x) , 
\label{eq_partondis}
\end{equation}
where $j^{q_V/p}_{bare}$ is the  valence quark distribution in the bare proton described by \( 3q\)  Fock states, 
and $\delta j^{q/p}$ includes both valence and sea contribution coming from the $BM$ Fock component of the proton state, i.e. $q=(u_V+\bar u, d_V+\bar d)$.
As we consider only the minimal $3q$ and $q\bar q$ configurations for the baryon and meson components in the $BM$ fluctuation, respectively, 
only the meson can contribute to the sea of the physical proton.

 \begin{figure}[h!]
\centering
\includegraphics[width=.8\textwidth]{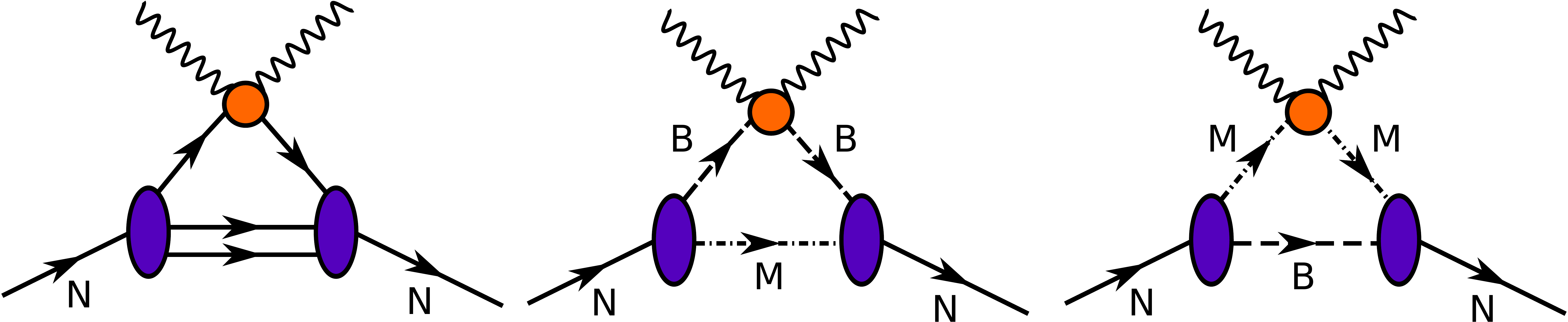}
\caption{Deeply virtual scattering from the bare nucleon (left figure), from the virtual baryon (middle figure, with the meson as spectator) and from the virtual meson (right figure, with the baryon  as spectator)  of the dressed nucleon.}
\label{fig_handbag}
\end{figure}

\noindent The last term in Eq.~\eqref{eq_partondis} can be further split into two contributions, with the active parton belonging either to the baryon ($\delta j^{q/BM}$) or to the meson ($\delta j^{q/MB})$, i.e.
 \begin{equation}
\delta j^{q} (x) =\sum_{B,M}\left[ \delta j^{q/BM}(x) + \delta j^{q/MB}(x) \right].
\label{eq:conv}
\end{equation}

\noindent The higher-Fock state contribution to the different parton distribution functions of the proton can be written as the following convolutions:\\

\noindent $\bullet$ for the unpolarized PDF:
\begin{eqnarray}
\delta f_{1}^{q/p}(x) &=& \sum_{B,M} \left [\int_x^1 \frac{dy}{y}\, f^{p/BM}(y)\, f_{1}^{q/B} \left(\frac{x}{y}\right)
   +\int_x^1 \frac{dy}{y}\,f^{p/MB}(y)\, f_{1}^{q/M} \left(\frac{x}{y}\right)
 \right ] \nonumber\\
 &&+\sum_{B,V} \int_x^1 \frac{dy}{y} f^{p/V B}_{LL} (y) f_{1LL}\left(\frac{x}{y}\right) , 
 \label{eq:splittingf1}
\end{eqnarray}
where the  sum over $B$ involves baryons of spin $1/2$,  $M$ stands for both scalar and vector mesons, while  $V$ refers to the contribution of only vector mesons.
In Eq.~\eqref{eq:splittingf1}
 the splitting functions are given by
\begin{eqnarray}
f^{p/BM}(y) = f^{p/MB}(1-y) 
&=& \int \frac{d^2\bf{k}_\perp}{2(2\pi)^3}\sum_{\lambda^\prime,\lambda^{\prime \prime}} 
\left\vert\phi^{+(p/BM)}_{\lambda^\prime \lambda^{\prime \prime}} (y,\bf{k}_\perp)\right\vert^2,\\
 f^{p/V B}_{LL}(y) &=& \sum_{\lambda_B} \Big[ -\frac{1}{3} \Big|  \phi^{+(p,BV)}_{ \lambda_B 1} (1-y,-{\bf k}_\perp) \Big|^2  +\frac{2}{3} \Big| \phi^{+(p,BV)}_{ \lambda_B 0} (1-y,-{\bf k}_\perp) \Big|^2 \\
 &&- \frac{1}{3}\Big|  \phi^{+(p,BV)}_{ \lambda_B -1} (1-y,-{\bf k}_\perp) \Big|^2 \Big]. \nonumber
\label{splitting:f1}
\end{eqnarray}
\noindent The description of a nucleon as a sum of \( BM \) Fock components is independent of whether the photon couples to the baryon or to the meson, so on general grounds the relation \( f^{N/BM}(y) =f^{N/MB}(1-y) \) must hold.
It simple means that when a baryon, which carries a momentum fraction \( y \), is struck by the photon, the remaining meson carries a momentum fraction \( 1 - y \).
Furthermore, this relation ensures charge conservation and momentum conservation automatically.\\

\noindent $\bullet$ for the longitudinally polarized PDF:
\begin{eqnarray}
\delta g_{1}^{q/p}(x) &=& \sum_{B,M} \int_x^1 \frac{dy}{y}\, \Delta_L f^{p/BM}(y)\, g_{1}^{q/B} \left(\frac{x}{y}\right)
   +\sum_{B,V}\int_x^1 \frac{dy}{y}\,\Delta_L f^{p/V B}(y)\, g_{1}^{q/V} \left(\frac{x}{y}\right),\label{eq:splittingg1}
\end{eqnarray}
with the splitting functions 
\begin{eqnarray}
\Delta_L f^{p/B M }(y) 
&=& \int \frac{d^2\bf{k}_\perp}{2(2\pi)^3} \sum_{\lambda^\prime \lambda^{\prime \prime}}
(-1)^{\frac{1}{2}-\lambda^\prime}
\left\vert\phi^{+(p/B M)}_{\lambda^\prime \lambda^{\prime \prime}} (y,\bf{k}_\perp)\right\vert^2,\\
\Delta_L f^{p/V B}(y) 
&=& \int\frac{d^2\bf{k}_\perp}{2(2\pi)^3} \sum_{\lambda^\prime} \left[
\left\vert\phi^{+(p/BV)}_{\lambda^\prime +1} (1-y,-\bf{k}_\perp)\right\vert^2
-
\left\vert\phi^{+ (p/BV)}_{\lambda^\prime-1} (1-y,-\bf{k}_\perp)\right\vert^2 \right].
\end{eqnarray}

\noindent $\bullet$ for the transversity:
\begin{eqnarray}
\delta h_{1}^{q/p}(x) &=& \sum_{B,M} \int_x^1 \frac{dy}{y}\, \Delta_T f^{p/BM}(y)\, h_{1}^{q/B} \left(\frac{x}{y}\right)
   +\sum_{B,V}\int_x^1 \frac{dy}{y}\,\Delta_T f^{p/V B}(y)\, \sqrt{2}h_{1}^{q/V} \left(\frac{x}{y}\right)
  ,\label{eq:splittingh1}
\end{eqnarray}
with the splitting functions 
\begin{eqnarray}
\Delta_T f^{p/BM}(y) &=& \int \frac{d^2\bf{k}_\perp}{2(2\pi)^3}
\sum_{\lambda^{\prime \prime}}\left[
\phi^{-(p/BM)}_{-\lambda^{\prime \prime}} (y,{\bf k}_\perp)\left(\phi^{+ (p/BM)}_{+\lambda^{\prime \prime}} (y,{\bf k}_\perp)\right)^*\right],\\
\Delta_T f^{p/V B}(y) &=& \int \frac{d^2\bf{k}_\perp}{2(2\pi)^3}
\sum_{\lambda^\prime}
\left[
\phi^{-(p/BV)}_{\lambda^\prime 0} (1-y,-{\bf k}_\perp)\left(\phi^{+ (p/BV)}_{\lambda^\prime +1} (1-y,-{\bf k}_\perp)\right)^*\right].
\end{eqnarray}
We note that in the convolution model an active meson (photon couples to the meson in the \( BM \) Fock component) with spin zero can only contribute to the unpolarized quark PDF \( f_1 \). 

\section{Modeling and LFWF overlap representation of the PDFs}
\label{sect:4}

In this section we specify the ingredients of the model calculation for the PDFs in the meson-cloud model.

The lowest-mass fluctuations for the proton which we include in our calculations are
\begin{eqnarray} 
     p (uud) &\rightarrow  & n(udd) \,\pi^+(u\bar d),\nonumber\\ 
     p (uud) &\rightarrow  & p(uud)\,  
 \pi^0 \left( \frac{1}{\sqrt{2}} [d\bar d - u\bar u]\right),\nonumber\\ 
     p (uud) &\rightarrow  & n(udd) \,\rho^+(u\bar d),\nonumber\\ 
  p (uud) &\rightarrow  &p(uud)\,  
 \rho^0 \left( \frac{1}{\sqrt{2}} [d\bar d - u\bar u]\right).
 \end{eqnarray}

For the vertex function we use the results which have been explicitly derived in Refs.~\cite{Pasquini:2007iz,Pasquini:2006dv}.
These results were obtained using TOPT in the infinite-momentum frame. In TOPT the intermediate particles are on their mass-shell.
However, an additional off-shell dependence is introduced in the vertex function for a vector meson due to the derivative coupling.
So, even using TOPT, we have a freedom on how to choose the meson energy in the vertex. 
In principle, there are two possible prescriptions:
\begin{itemize}
 \item[A)] $p^\mu_V=(E_V,\mathbf p_V)$, with the on-shell meson energy $E_V=\sqrt{m_{V}^{2}+\mathbf p_{V}^{2}}$,
 \item[B)]$p^\mu_V=(E_V,\mathbf p_V)$, with the off-shell meson energy $E_V=E_N-E_B$.
\end{itemize}
We will adopt the choice B), following the arguments of  Ref.~\cite{Bissey:2005kd} to establish a correspondence between time-ordered perturbation theory in the infinite momentum frame and light-front perturbation theory.
Furthermore, 
because of the extended structure of the hadrons involved, one has also 
to multiply the coupling constant 
for pointlike particles in the vertex function by phenomenological vertex form factors.
These form factors  parametrize the unknown 
microscopic effects at the vertex and have to obey the constraint
$F_{NBM}(y,k_\perp^2 )=F_{NBM}(1-y,k_\perp^2 )$ to ensure basic properties 
like charge and momentum conservation simultaneously~\cite{Melnitchouk:1998rv}.
To this aim we will use the following functional form 
\begin{equation} 
F_{NBM}(y,k_\perp^2 )= \mbox{exp}
\left[\frac{M_N^2-M^2_{BM}(y,\bold{k}_\perp)}{2\Lambda_{BM}^2}\right],
\label{eq:74}
\end{equation}
where $\Lambda_{BM}$ is a cut-off parameter.
Following the recent analysis of Refs.~\cite{Feng:2012gu,Traini:2013zqa}, we take $\Lambda_{BM}=0.8$ GeV for all the baryon-meson fluctuation entering into our calculation.
For the $NBM$
coupling constants at the interaction vertex, we used the
numerical values given in Refs.~\cite{Machleidt:1987hj,Holtmann:1996be},  
i.e. $g^2_{NN\pi}/4\pi=13.6$, $g^2_{NN\rho}/4\pi=0.84$  and $f_{NN\rho}=6.1 g_{NN\rho}$\footnote{Note that  we follow Ref.~\cite{Pasquini:2007iz} for the vertex interaction, where  the coupling constant $f_{NN\rho}$ is dimensionless. In order to compare with the definition adopted in Refs.~\cite{Cao:2003zm}, $f_{NN\rho}$ has to be multiplied by a factor $4M_N$.}.

\noindent With this choice of the parameters, 
in the case of the 
$p\rightarrow$ $N\pi$ and $p\rightarrow$ $N\rho$ transitions, one finds
\begin{eqnarray}
& & P_{p/N \pi }=P_{p/p \pi^0 }+P_{p/n\pi^+ }=3P_{p/p \pi^0 }= 13.17\%,\\
& & P_{p/N \rho }=P_{p/p \rho^0 }+P_{p/n\rho^+ }=3P_{p/p \rho^0 }= 2.17 \%.
\end{eqnarray}

 For the hadron states of the bare nucleon and baryon-meson components in 
Eq.~(\ref{eq_physnuc_fock_state}) we adopt the light-front constituent quark model of Ref.~\cite{Pasquini:2007iz}, that we briefly summarize here for convenience.
\\
A hadron state with momentum  \( \tilde{p} \) and helicity \( \Lambda \) is given by
\be
\ket{{\tilde{p}_H},\Lambda;H} = \sum_{q_i,\lambda_i}
\int\left[\frac{{\rm d}x }{\sqrt{x}}\right]_N
[{\rm d}^2{\mathbf k}_\perp]_N
\Psi^{[H],\Lambda; q_1\dots q_N}_{\lambda_1\dots \lambda_N}(\{\tilde k_i\}_{i=1,...,N})
\prod_{i=1}^N
\ket{x_i p^+_H, \, {\mathbf p}_{i\perp},\lambda_i,q_i},
\label{eq_hadronic_state}
\ee
where \( \Psi^{[H],\Lambda; q_1\dots q_N}_{ \lambda_1 \dots \lambda_N }\left(\{x_i,\mathbf{k}_{i \perp} \} \right) \) is the LFWF which gives the probability amplitude for finding in the hadron with a light-front helicity \( \Lambda \), \( N \) partons with momenta \( (x_i p^+_H, {\mathbf p}_{i\perp}=\,{\mathbf k}_{i\perp} + x_i {\mathbf p}_{\perp H}) \), with \( x_i \) being the momentum fraction of the \( i\)-th parton (the index \( i \) runs from \( 1 \) to \( N \)) w.r.t. its parent hadron and \( \mathbf{k}_{i \perp}\) being its intrinsic transverse momentum. The index \( \lambda_i \) labels the helicity and \( q_i \) the isospin of the \( i \)-th parton, respectively.
In Eq.~(\ref{eq_hadronic_state}) and in the following, the integration measures are defined by
\begin{eqnarray}
\left[\frac{{\rm d} x}{\sqrt{x}}\right]_N 
&=& \left(\prod_{i=1}^N \frac{{\rm d} x_i}{\sqrt{x_i}}\right)
\delta\left(1-\sum_{i=1}^N x_i\right),\\ 
\left[ {\rm d}^2{\mathbf k}_\perp\right]_N &=& \frac{1}{\left( 2(2\pi)^3 \right)^{N-1}}\left(\prod_{i=1}^N
{\rm d}^2{\mathbf k}_{\perp\,i}\right)\, 
\delta\left(\sum_{i=1}^N {\mathbf k}_{\perp\,i}\right).
\label{eq_def_integration_measures}
\end{eqnarray}
By taking into account the minimal Fock-state component, one has $N=3$ and $N=2$ for the baryon and meson, respectively.

\noindent As explained in Ref.~\cite{Boffi:2002yy}, the wave function \( \Phi^{[H],\Lambda; q_1\dots q_N}_{\lambda_1 \dots \lambda_N} \) can be obtained by transforming the ordinary equal-time (instant-form) wave function  into that in the light-front dynamics.
The instant-form wave function\( \Phi^{[H],\Lambda; q_1\dots q_N}_{\mu_1 \dots \mu_N} \) is constructed as the product of a momentum wave function \( \tilde{\psi}^{[H]} \left( \{ \tilde{k}_i \}\right) \), which is spherically symmetric and invariant under permutations, and a spin-isospin wave function \( \phi^{[H]} \left( \{ \mu_i \}, \{ q_i \} \right) \), which is uniquely determined by \( SU(6)\)-symmetry requirements, i.e.
\begin{equation}
 \Phi^{[H],\Lambda; q_1\dots q_N}_{\mu_1 \dots \mu_N} \left( \{ \tilde{k}_i \}\right)= \tilde{\psi}^{[H]} \left( \{ \tilde{k}_i \}\right) \otimes  \phi^{[H]} \left( \{ \mu_i \}, \{ q_i \} \right),
\end{equation}
\noindent where $\mu_i$ is the canonical (instant-form)  helicity of the \(i\)-th parton.
\noindent The transformation to the light-front form can be obtained by taking  into account relativistic effects such as the Melosh-Wigner rotation
\begin{equation} 
\Psi^{[H],\Lambda;\{q_i \}}_{\{ \lambda_i \}}\left(\{x_i,\mathbf{k}_{\perp i} \} \right) = \tilde{\psi}^{[H]} \left( \{ \tilde{k}_i \}\right) \sum_{\mu_1,\dots,\mu_N}  \phi^{[H]} \left( \{ \mu_i \}, \{ q_i \} \right) \prod_{i=1}^N {D}^{1/2\,*}_{\mu_i\lambda_i}(R_{cf}({ \tilde k}_i)),
\end{equation}
where ${D}^{1/2\,*}_{\mu_i\lambda_i}(R_{cf}({\tilde k}_i))$ are the Melosh
 rotations defined in Ref.~\cite{Boffi:2002yy}.\\

In the case of the nucleon, we consider two different models for the momentum wave function.
The first one (hereafter referred as \textit{model 1}) is based on a phenomenological Ansatz for the momentum dependence 
of the light-front wave function that has been originally assumed to fit the electroweak form factors~\cite{Schlumpf:94a,Schlumpf:94b}.
Recently, it has been  been applied for the calculation of a variety of parton distributions ~\cite{Lorce:2011dv}, including  leading-twist  GPDs  and collinear PDFs~\cite{Boffi:2002yy,Boffi:2003yj,Pasquini:2004gc,Pasquini:2005dk,Pasquini:2006iv,Boffi:2007yc},  leading- and higher-twist transverse momentum dependent parton distributions (TMDs)~\cite{Pasquini:2008ax,Boffi:2009sh,Pasquini:2010af,Pasquini:2011tk,Lorce:2014hxa}, and electroweak form-factors~\cite{Pasquini:2007iz}.
The momentum wave function is given by
\bea
\label{eq:psifc}
\tilde{\psi}^{[N]}(\{\tilde k_i\}_{i=1,2,3})
=2(2\pi)^3\left[\frac{1}{M_0}\frac{\omega_1\omega_2\omega_3}{x_1x_2x_3}\right]^{1/2}\frac{N'}{(M_0^2+\beta^2)^\gamma},
\label{eq_schlumpf_wave_function}
\eea 
where  $\omega_i=\sqrt{m^{2}_{q}+{\mathbf k}_i^2}$ is the energy
of the $i$-th quark, $M_0=\sum_i \,\omega_i$ is the free mass
of the system of $N$ non-interacting quarks and
 $N'$  a normalization factor such that $\int d[x]_3 d[\mathbf k_{\perp}]_3|\psi(\{\tilde k_i\}_{i=1,2,3}) |^2=1$. 
 In Eq.~(\ref{eq_schlumpf_wave_function}), the scale $\beta$,
the parameter $\gamma$ for the power-law behavior, and the quark mass $m_q$ are 
taken from the fit to the nucleon electroweak form factor in the light-front meson-cloud model of Ref.~\cite{Pasquini:2007iz}, i.e. $\gamma= 3.21$, $\beta= 0.489$ GeV,
and $m_q=0.264$ GeV. 

As alternative model (hereafter referred as \textit{model 2}), we will discuss the predictions obtained within a relativistic hypercentral quark model~\cite{Faccioli:1998aq}, which extends the non-relativistic version of Ref.~\cite{Ferraris:1995ui} and has been recently applied within a light-front meson-cloud model for the unpolarized PDF~\cite{Traini:2011tc,Traini:2013zqa}.
The hypercentral model is based on the mass operator $M=M_0+V$, with the interaction given by
\bea
V=\frac{\tau}{y}+\kappa_l y,
\eea
where $y=\sqrt{\rho^2+\lambda^2}$ is the radius of the hypersphere in six dimensions, and $\rho$ and $\lambda$ are the intrinsic Jacobi coordinates $\boldsymbol{\rho}=(\mathbf r_1-\mathbf r_2)/\sqrt{2}$ and $\boldsymbol{\lambda}=(\mathbf r_1+\mathbf r_2-2\mathbf r_3)/\sqrt{6}$.
The model depends on two parameters, $\tau$ and $\kappa_l$, which have been fixed to reproduce the basic features of the low-lying nucleon spectrum ($\tau=3.3$ and $\kappa_l=1.80$ fm$^{-2}$, from Refs.~\cite{Traini:2013zqa,Faccioli:1998aq}).
The momentum-dependent wave function is taken with orbital angular momentum $L=0$, and reads
 \bea
\label{eq:psifchyp}
\tilde{\psi}^{[N]}(\{\tilde k_i\}_{i=1,2,3})
=2(2\pi)^3\left[\frac{1}{M_0}\frac{\omega_1\omega_2\omega_3}{x_1x_2x_3}\right]^{1/2}
\psi_{0,0}(y){\cal Y}^{(0,0)}_{[0,0,0]}(\Omega),
\eea
where $\psi_{\gamma,\nu}(y)$ is the hyperradial wave function solution of the eigenvalue problem for the mass operator $M$, which is expanded on a truncated 
set of hyper-harmonic oscillator basis states, and ${\cal Y}^{(L,M)}_{[\gamma,l_\rho,l_\lambda]}(\Omega)$ are the hyperspherical harmonics defined on the hypersphere of radius one.
  
\noindent 
For the pion, we choose the LWFW proposed in Refs.~\cite{Schlumpf:1994bc,Chung:1988mu}, which has been applied to calculate GPDs~\cite{Frederico:2009fk}, and leading- and higher-twist TMDs~\cite{Pasquini:2014ppa,Lorce:2016ugb}.
The explicit expression for the momentum dependent part of the LFWF reads \begin{eqnarray}
\tilde{\psi}^{[\pi]}(\bar x,\mathbf k_\perp)
=[2(2\pi)^3]^{1/2}\left[\frac{M_0}{4\bar x(1-\bar x)}\right]^{1/2}\frac{i}{\pi^{3/4}\alpha^{3/2}}
\exp{[-k^2/(2\alpha^2)]},
\label{eq:can_psi}
\end{eqnarray}
with $\mathbf k=\mathbf k_1=-\mathbf k_2$, $\bar x=x_1=1-x_2$, and the two parameters 
$\alpha=0.3659$ GeV and $m_q=0.22$ GeV from Ref.~\cite{Pasquini:2007iz}.
The phase of the pion wave function (\ref{eq:can_psi}) 
is consistent with that of the antiquark spinors of Ref.~\cite{Brodsky:1989pv}.

\noindent
The wave function of the $\rho$ differs from the pion only in the spin 
component, with the canonical spin states of the $q\bar q$ pair coupled to $J=1$ instead of $J=0$.\\

The light-front formalism allows us to obtain a convenient representation of the hadron PDFs in terms of overlap of LFWFs.
Choosing to label the active quark with $i=1$, the hadron light-front helicity amplitudes introduced in App.~\ref{appendix:a} can be obtained 
as

\begin{eqnarray}
A^{q/H}_{\Lambda^\prime\lambda^\prime,\Lambda\lambda}=\int d[1\dots N]
\sum_{\lambda_2,\dots,\lambda_N}\sum_{q_1\dots q_N}
\left(\psi^{[H],\Lambda^\prime;q_1\dots q_N}_{\lambda^\prime\lambda_2\dots\lambda_N}\right)^*
\psi^{[H],\Lambda;q_1\dots q_N}_{\lambda\lambda_2\dots\lambda_N}.
\end{eqnarray}

\noindent For $N=3$
\begin{eqnarray}
d[12 3]=[dx]_3[d^2\mathbf k_\perp]_3 \, 3 \, \delta(x-x_1),
\end{eqnarray}
and for $N=2$
\begin{eqnarray}
d[12]=[dx]_2[d^2\mathbf k_\perp]_2  \, \delta(x-x_1).
\end{eqnarray}

\noindent From the relations in \eqref{amplPDFs1/2}, we then find the following LFWF overlap representation for the contribution of the $3q$ Fock-state to the proton PDFs
\begin{subequations}
\begin{align}
f^{q/p}_1&=\int\ud[123] \, \sum_{\lambda_2\lambda_3}  \sum_{q_2q_3} \, \left[\Big|\Psi^{[p]+;qq_2q_3}_{+\lambda_2\lambda_3}\Big|^2 + \Big |\Psi^{[p]+;qq_2q_3}_{-\lambda_2\lambda_3}\Big|^2\right],\\
g^{q/p}_{1}&=\int\ud[123] \, \sum_{\lambda_2\lambda_3}\sum_{q_2q_3} \, \left[\Big |\Psi^{[p]+;qq_2q_3}_{+\lambda_2\lambda_3}\Big|^2-\Big|\Psi^{[p]+;qq_2q_3}_{-\lambda_2\lambda_3}\Big|^2\right],\\
h^{q/p}_1&=\int\ud[123] \, \sum_{\lambda_2\lambda_3}\sum_{q_2q_3} \, \left(\Psi^{[p]+;qq_2q_3}_{+\lambda_2\lambda_3}\right)^*\Psi^{[p]-;qq_2q_3}_{-\lambda_2\lambda_3}.
\end{align}
\end{subequations}

\noindent Analogously, the contribution of the \( q\bar{q} \) Fock-state to the pion PDF reads
\begin{subequations}
\begin{align}
f^{q/\pi}_1(x)&=f^{\bar q/\pi}_1(x)=\int\ud[12]\, \sum_{\lambda_2} \, \left[ \Big |\Psi^{[\pi]; q\bar q}_{+\lambda_2}\Big|^2 + \Big|\Psi^{[\pi];q\bar q}_{-\lambda_2}\Big|^2\right],
\end{align}
\end{subequations}
where $ f^{q/\pi}=f^{q/\pi^+}$ refers to the parton distribution in the charged pion $\pi^+$, while  the other PDFs can be obtained by isospin symmetry and charge symmetry, i.e. 
$f^{u/\pi^+}=f^{\bar d/\pi^+}=f^{d/\pi^-}=f^{\bar u/\pi^-}=2f^{u/\pi^0}=2f^{\bar u/\pi^0}=2f^{d/\pi^0}=2f^{\bar d/\pi^0}$.

\noindent Using the relations in \eqref{amplPDFs1} for the vector meson, we also obtain the following LFWF overlap representation for the contribution of the $q\bar q$ Fock-state to the PDFs of the $\rho$ meson
\begin{subequations}
\begin{align}
f^{q/\rho}_1(x)&=f^{\bar q/\rho}_1(x)=\frac{2}{3}\int\ud[12]\, \sum_{\lambda_2}\left[\Big |\Psi^{[\rho] 0;q\bar q}_{+\lambda_2}\Big|^2+
\Big|\Psi^{[\rho]+1;q\bar q}_{+\lambda_2}\Big|^2+\Big|\Psi^{[\rho]-1;q\bar q}_{-\lambda_2}\Big|^2\right],\\
f^{q/\rho}_{1LL}(x)&=f^{\bar q/\rho}_{1LL}(x)=\int\ud[12] \, \sum_{\lambda_2}\left[2 \Big |\Psi^{[\rho] 0;q\bar q}_{+\lambda_2}\Big|^2-
\Big|\Psi^{[\rho]+1;q\bar q}_{+\lambda_2}\Big|^2-\Big|\Psi^{[\rho]-1;q\bar q}_{-\lambda_2}\Big|^2\right],\\
g^{q/\rho}_{1}(x)&=g^{\bar q/\rho}_{1}(x)=\int\ud[12] \, \sum_{\lambda_2}\left[\Big|\Psi^{[\rho] +1;q\bar q}_{+\lambda_2}\Big|^2-
\Big |\Psi^{[\rho]-1;q\bar q}_{-\lambda_2}\Big|^2\right],\\
h^{q/\rho}_{1}(x)&=h^{\bar q/\rho}_{1}(x)=\frac{1}{\sqrt{2}}\int\ud[12] \, \sum_{\lambda_2}\left[\Psi^{[\rho] 0;q\bar q}_{-\lambda_2}\left(\Psi^{[\rho] +1;q\bar q}_{+\lambda_2}\right)^*
+\Psi^{[\rho] -1;q\bar q}_{-\lambda_2}\left(\Psi^{[\rho] 0;q\bar q}_{+\lambda_2}\right)^*
\right],\end{align}
\end{subequations}
where $ j^{q/\rho}=j^{q/\rho^+}$ refers to the generic PDF $j$ in the charged $\rho$ meson $\rho^+$, while  the other PDFs can be obtained by isospin symmetry, i.e. 
$j^{u/\rho^+}=j^{\bar d/\rho^+}=j^{d/\rho^-}=j^{\bar u/\rho^-}=2j^{u/\rho^0}=2j^{\bar u/\rho^0}=2j^{d/\rho^0}=2j^{\bar d/\rho^0}$.

\section{Results and Discussion}
\label{sect:5}

In this section we discuss the results from the light-front meson-cloud model for the leading-twist PDFs of the proton, in comparison with available experimental extractions for the valence quark and sea quark contribution as well as for the flavor asymmetries in the unpolarized and polarized sea. Furthermore we present predictions for the nucleon tensor charge in comparison with other model calculations and phenomenological extractions.

Parton distribution functions are defined within a certain regularization scheme at a given 
factorization scale. 
The results within the light-front meson-cloud model refer to an assumed initial scale $Q_{0}^{2}$, where the nucleon state is expanded in the Fock space  in terms of the minimum (valence) and next-to-minimum (an extra $q \bar q$ pair) components, as described in Sect.~\ref{sect:2}.
To determine the initial matching scale consistent with QCD evolution we follow a standard procedure, that we shortly review.

\noindent We restrict our discussion at leading order (LO) in perturbation theory, and we evolve back at LO
the unpolarized parton distributions until the  momentum fraction carried by the valence quark matches the value calculated in the model.
The  momentum fraction carried by the valence quark 
is obtained from the $N=2$ Mellin moment of the non-singlet (NS) combination of the unpolarized PDFs, $\langle q_{{\rm NS}}(Q^2)
\rangle_N=\sum_q \int{\rm d}x\, x^{N-1} (f^q-f^{\bar q})(x,Q^2)$, 
 which evolves at LO according to the following equation
\bea
\label{evolution-moment}
\frac{\langle q_{{\rm NS}} (Q^2)\rangle_N}{\langle q_{{\rm NS}}(Q^{2}_{0}\rangle_{N}}\Bigg|_{LO} = \left( \frac{\alpha_{s} (Q^2) }{\alpha_{s}(Q_0^2)}\right)^{\frac{P_{NS}^{(0)(N)}}{2\beta_0}}.
\eea
In Eq.~\eqref{evolution-moment}, $ \beta_0 = 11- \frac{2 N_f}{3} $ is the lowest expansion coefficient of the QCD beta function, and 
the $N$-moment of the LO-NS splitting function $P_{NS}^{(0)}(N)$ has the following expression
  \begin{equation}
  P_{NS}^{(0)}(N) = \frac{8}{3}\Big[ 1 -\frac{2}{N(N+1)} + 4 \sum_{j=2}^{N}\frac{1}{j} \Big].
 \end{equation}
 We work in the scheme of variable flavor number $N_f$, with heavy-quark mass thresholds
 $m_c= 1.4$ GeV, $m_b =4.75$ GeV, $m_t =175$ GeV and  the strong coupling constant 
 $\alpha_s(M_{z}^{2})=0.13939$ corresponding to $\Lambda_{{\rm LO}}^{(3,4,5)}=359, \, 322,\,  255$ MeV.
Taking the phenomenological value $\langle q_{{\rm NS}}(Q^2)\rangle_2=0.35$ at $Q^2=10$ GeV$^2$ from the MSTW LO parametrization \cite{Martin:2009iq}, we reproduce the value of the model $\langle q_{{\rm NS}}(Q^2)\rangle_2=0.94$ at the scale $Q^{2}_0=0.186$ GeV$^2$.
The effect of introducing a non-perturbative quark sea carrying 6$\%$ of the nucleon momentum is to produce a higher input scale  with respect to the scenario with a bare nucleon 
consisting of only a 3$q$ valence component, where the hadronic scale of the model was found  
$\mu^2=0.176$ GeV$^2$~\cite{Pasquini:2011tk}.\\ 
\begin{figure}[t]
 \includegraphics[width=0.48\textwidth]{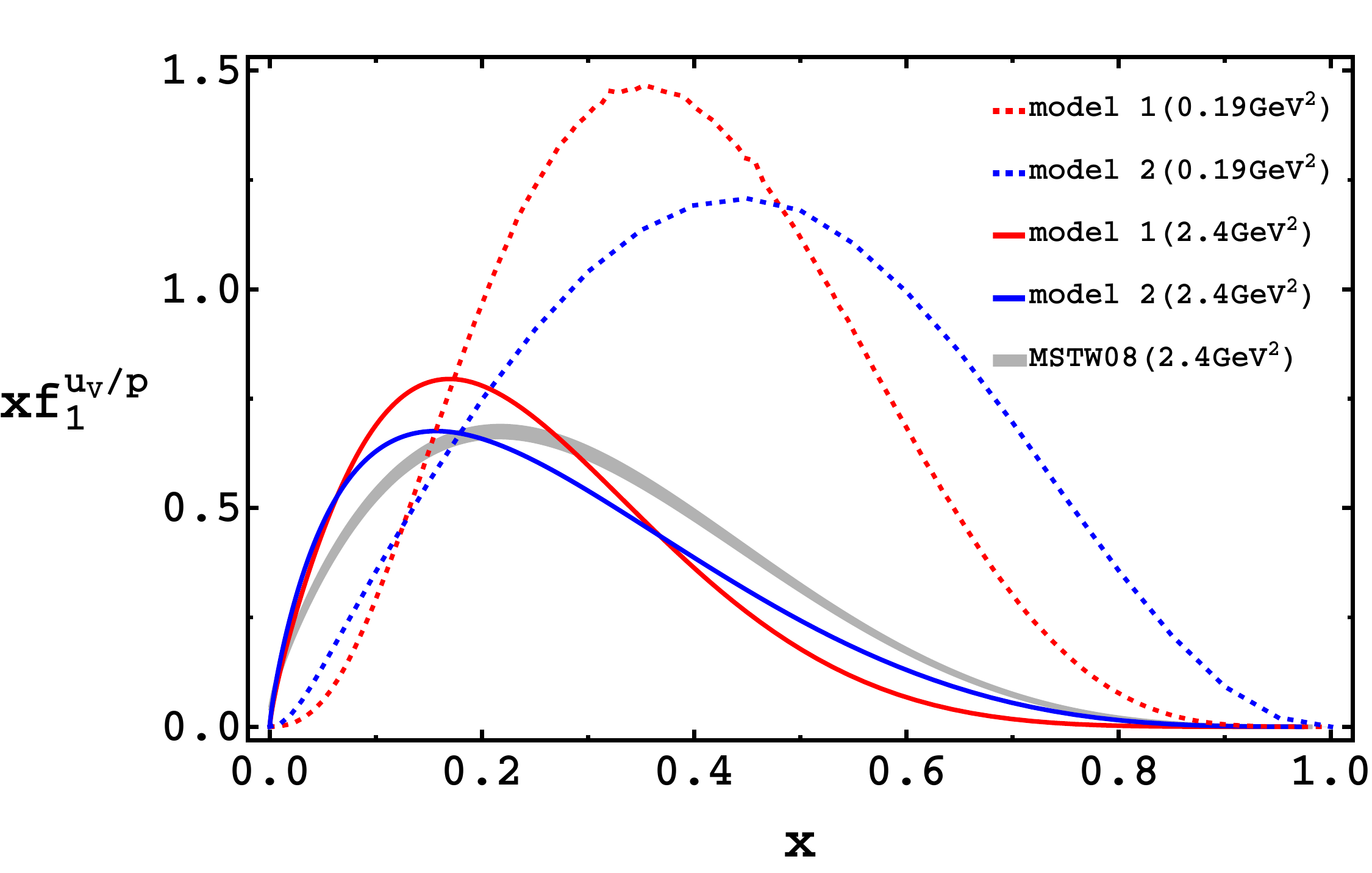}
  \hspace*{0.2cm}
  \includegraphics[width=0.48\textwidth]{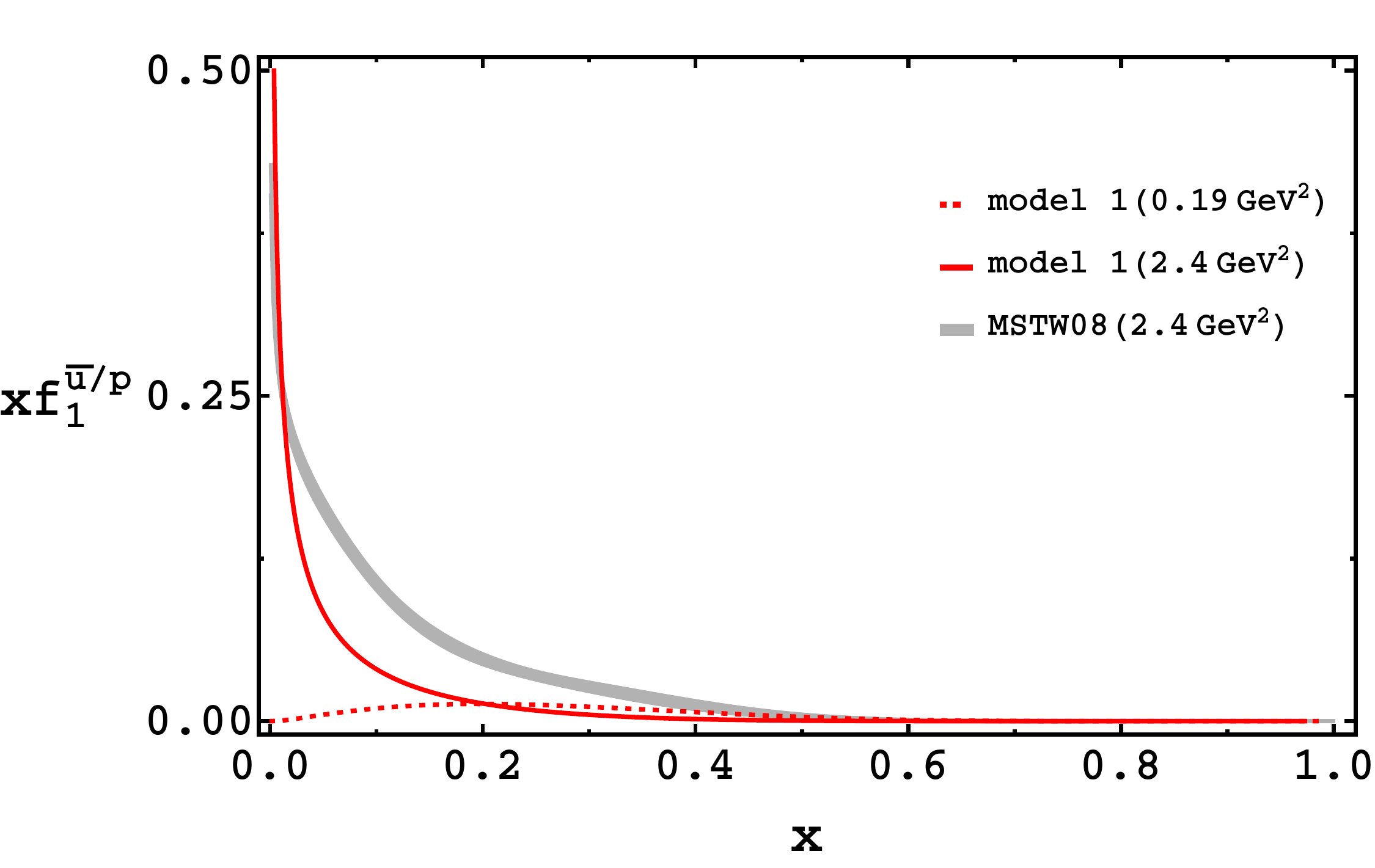} \includegraphics[width=0.48\textwidth]{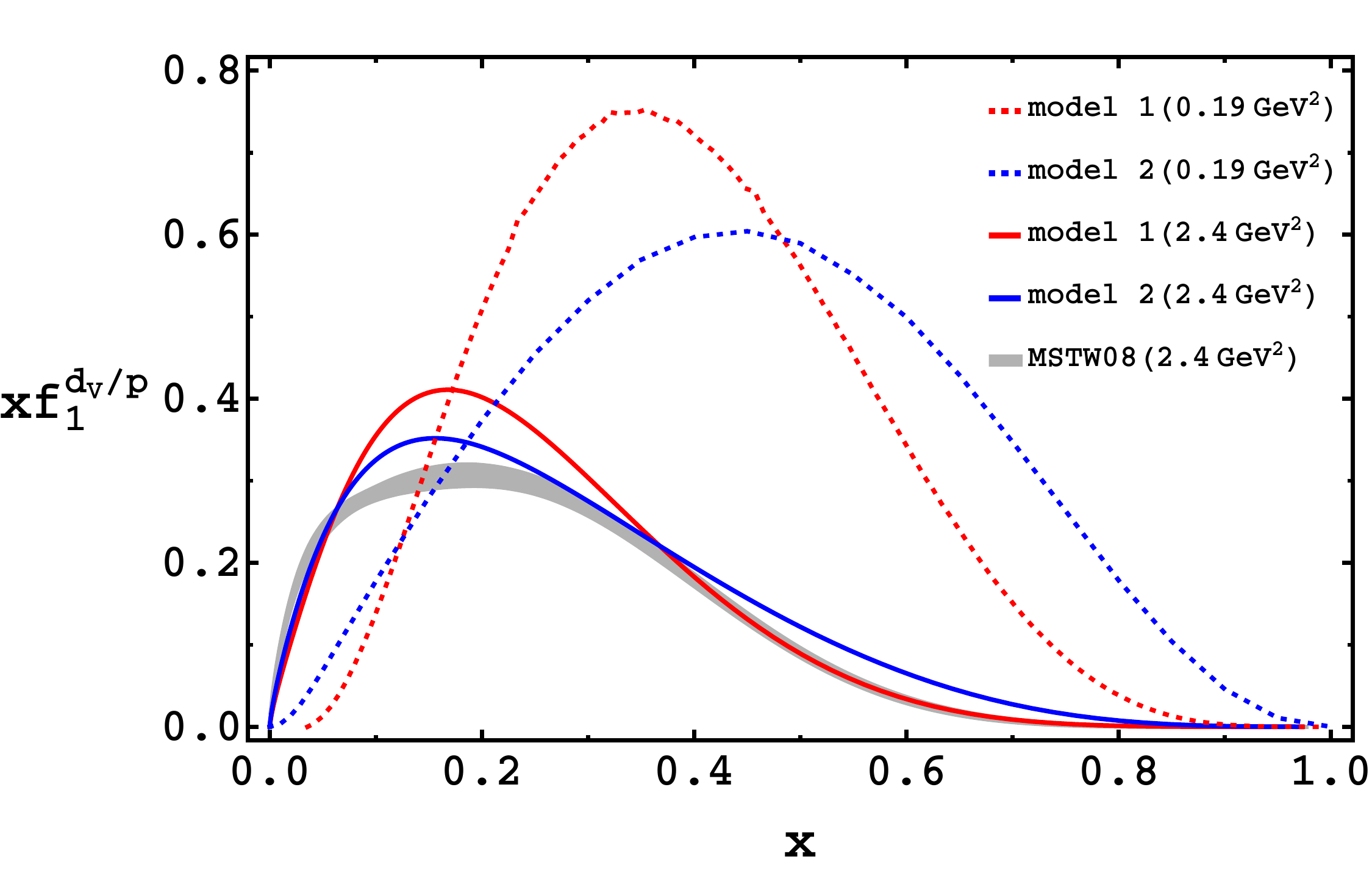}
  \hspace*{0.2cm}
  \includegraphics[width=0.48\textwidth]{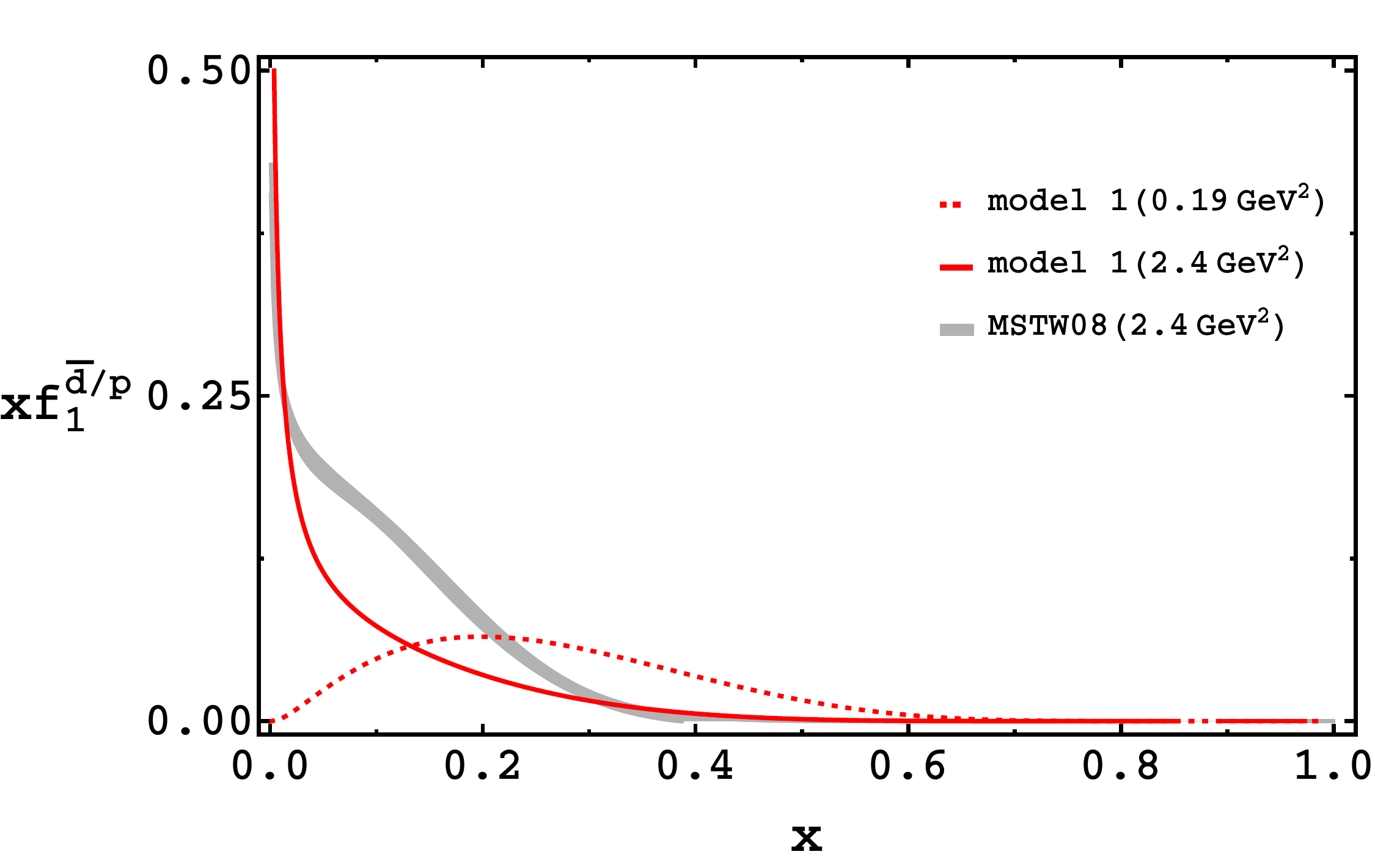}
  \caption{The unpolarized parton distribution $x f_1$ for the up (upper panels) and down (lower panels) quark as function of $x$.
  The left and right panel show the results for the valence and sea quark contributions, respectively.
The red and blue curves are obtained  within the light-front meson-cloud model from the \textit{model 1} and \textit{2} for the nucleon LFWF, respectively, at the input scale $Q^{2}_{0} $  (dotted curves) and after LO evolution to 
  $Q^{2}=2.4 $ GeV$^2$  (solid curves).  
  In the case of the sea quark contribution the results from \textit{model 2} are indistinguishable from  \textit{model 1} and are not shown.
  The grey bands show the  phenomenological results from the MSTW08 parametrization of Ref.~\cite{Martin:2009iq} at $Q^{2}=2.4 $ GeV$^2$.}
  \label{fig_f1_up_valence_sea}
\end{figure}
\begin{figure}[h!]
 \includegraphics[width=0.48\textwidth]{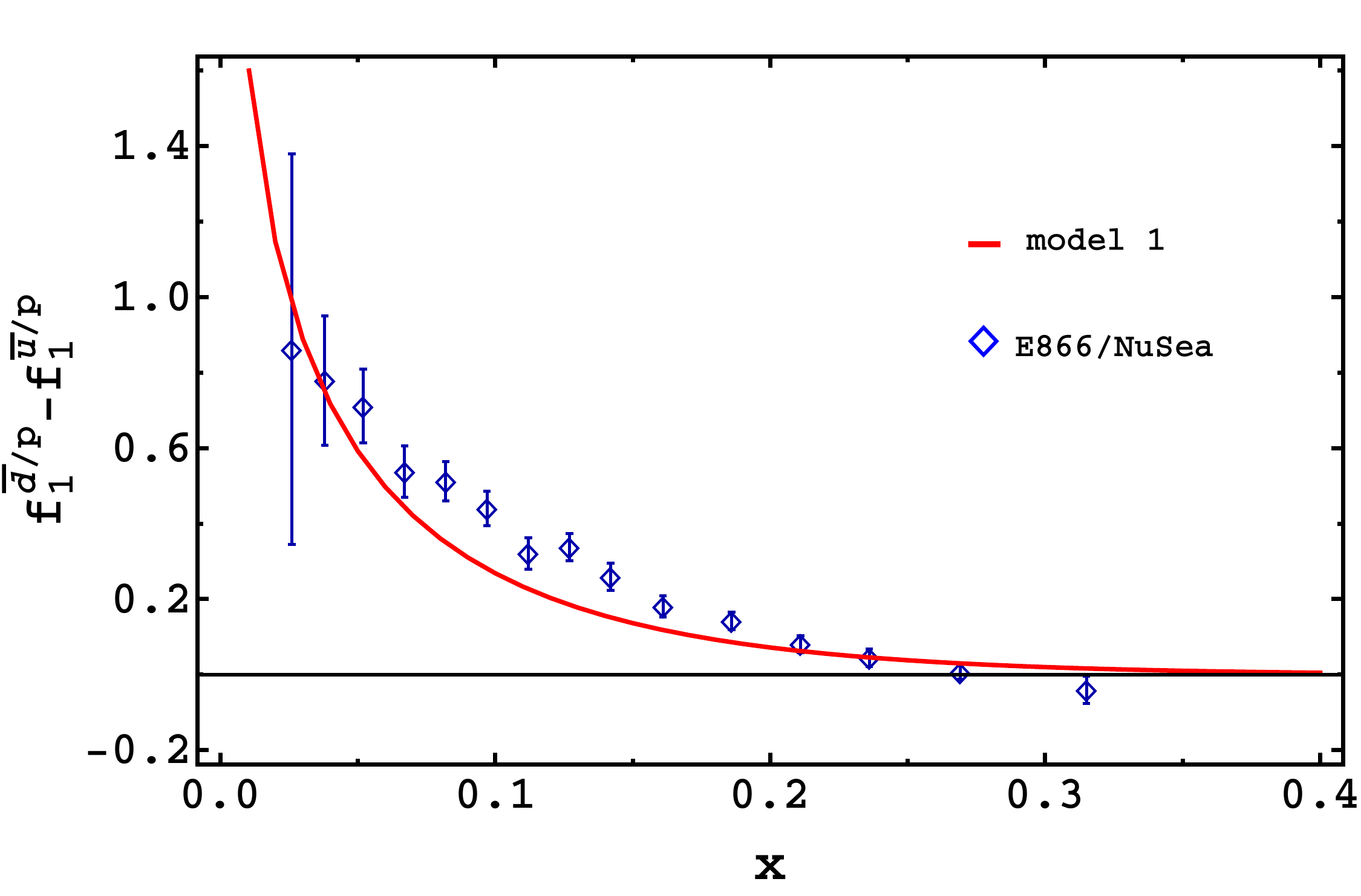}
 \hspace*{0.2cm}
  \includegraphics[width=0.48\textwidth]{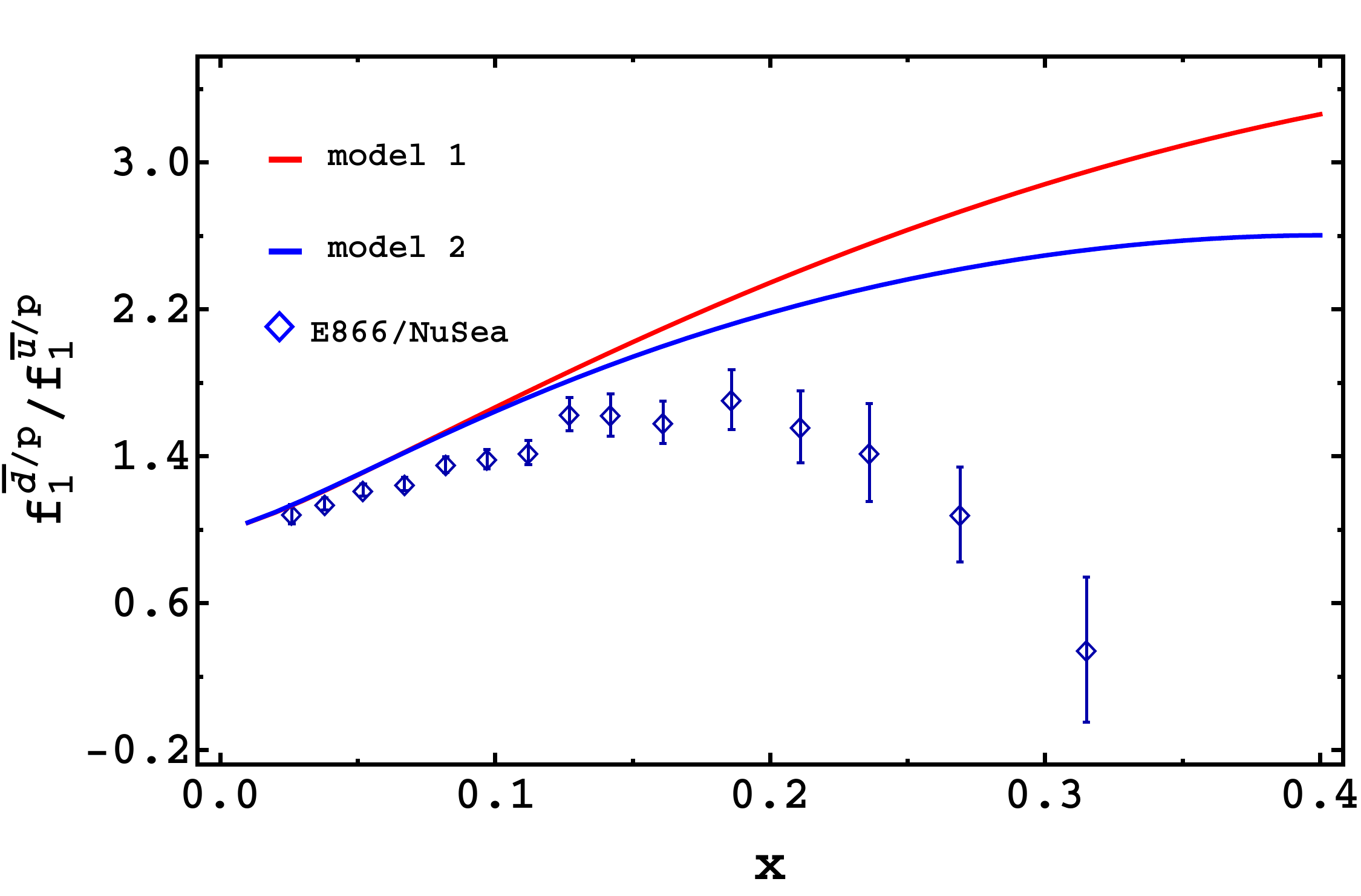}
  \caption{Result from the light-front meson-cloud model for the  flavor asymmetry of the sea distribution  \(f_1^{\bar{d}/p}(x) - f_1^{\bar{u}/p}(x) \)  (left panel) and the ratio \( f_1^{\bar{d}/p}(x) /  f_1^{\bar{u}/p}(x) \)  (right panel) as function of $x$, after LO evolution to  the scale \( Q^2= 54~\mbox{GeV}^2 \) of the E866 experiment. The red and blue curves are obtained with the \textit{model 1} and \textit{2} for the nucleon LFWF, respectively, and are indistinguishable in the case of the sea quark flavor asymmetry. The E866 experimental data  are from Ref.~\cite{Towell:2001nh}. }
  \label{fig_unpolarised_sea_asymmetry}
\end{figure}

In Fig.~\ref{fig_f1_up_valence_sea} we show the valence-quark and the sea-quark contribution to \( f_1\)  for  up and down quarks, using both the \textit{model 1} and \textit{model 2} for the bare nucleon LFWF.
The difference between the two model calculations for the valence contributions are more pronounced at the initial scale of the model $Q_{0}^{2}$ (dotted curves), and becomes smaller after evolution to the scale $Q^2=2.4$ GeV$^2$. The evolved results are overall in reasonable agreement with the MSTW parametrization at $Q^2=2.4$ GeV$^2$. However, in the case of the up-quark distribution, we notice that both models have a peak position at lower value of $x$ with respect to the phenomenological parametrization, while the fall-off at higher $x$ is better reproduced in \textit{model 2}.
In the case of the distribution for down quark, the peak position is well reproduced in both models 
with a very good agreement between \textit{model 1} and the phenomenological parametrization for $x> 0.3$.
\\
The sea-quark distributions are the same in the two models at the input scale, as they are generated by the antiquarks of the $\pi$ and $\rho$ in the $N\rightarrow NM$ fluctuations. 
After LO evolution, the difference for the sea-quark unpolarized distributions in the two models turns out to be so small that is practically indistinguishable in the plot. Therefore, we decided to show only the result for \textit{model 1}.
The perturbative evolution plays an important role at $x<0.1$ once the non-perturbative content is introduced at the hadronic scale $Q_{0}^2$, while for both flavors the distributions miss strength at intermediate values of $x$  with respect to the phenomenological parametrization.
However, the Sullivan process  is one of the most successful non-perturbative mechanism in explaining the flavor asymmetry of the unpolarized sea (see Refs.~\cite{Speth:1996pz,Londergan:1998ai,Chang:2014jba,Kumano:1997cy} for comprehensive reviews about different applications  of the meson-cloud model).

\noindent In Fig.~\ref{fig_unpolarised_sea_asymmetry} we show our results after LO evolution  for the sea-quark unpolarized flavor asymmetry \( f_1^{\bar{d}/p}(x) - f_1^{\bar{u}/p}(x) \) and the ratio \( f_1^{\bar{d}/p}(x) /  f_1^{\bar{u}/p}(x) \) in comparison with the experimental data from the E866 experiment at $Q^2=54$ GeV$^2$~\cite{Towell:2001nh}.
It is well known that a QCD evolution at LO with an SU(6) symmetric input can not generate an asymmetric sea.
The excess of $\bar d$  in the light-front meson-cloud model at the input scale is 
responsible for the observed asymmetry, and it is in good agreement with the experimental data after evolution to  the relevant experimental scale.
More sophisticated meson-cloud models including also the contribution from $N\rightarrow \Delta \pi$ fluctuations have been discussed in literature~\cite{Melnitchouk:1998rv,Kumano:1990mj,Kumano:1990mm,Alberg:2012wr,Salamu:2014pka,Traini:2013zqa}. Owing to the freedom in the choice 
of the cutoff in the  baryon form factors at the $NBM$ vertices,  the contributions of opposite sign from the $N\pi$ and $\Delta \pi$ fluctuations accommodates 
to produce very similar results as in our approach.  Furthermore,  the $N\Delta \pi$ 
fluctuations do not contribute to the polarized quark sea contribution, and therefore they will not  further be discussed in the following.
 It is also encouraging to observe that our treatment with perturbative evolution at LO produces very similar results of more complex calculations at NLO and NNLO~\cite{Traini:2013zqa}. 
 The results  for the ratio of the $\bar d$ over the $\bar u$ contribution to $f_1$ do not reproduce the rapid decrease  of the data towards and below unity for  $x>0.2 $. 
 This finding is common to different  calculations within meson-cloud models and  non-perturbative models including chiral perturbation theory and instanton models~\cite{Peng:1998pa,Alberg:1999bc,Nikolaev:1998se,Szczurek:1994spe,Pobylitsa:1998tk,Dorokhov:1993fc}. On the other side, among the most recent sets of parton distribution fits, CT14~\cite{Dulat:2015mca} and MMHT14~\cite{Harland-Lang:2014zoa}  reproduce this trend of the data, at variance with  the PDF fit provided by the statistical  model~\cite{Basso:2015lua} that predicts a ratio larger than one at larger $x$.
However, as $x$ increases beyond $0.25$, the data become less precise. The  new Drell-Yan measurements  of the Fermilab E906/SeaQuest experiment will help understanding this region better~\cite{Reimer:2011zza}. 
By now, only preliminary results from the 2015 data set of the E906 experiment have been shown at conferences~\cite{Lorenzon}, and support the predictions of a ratio larger than one at larger $x$.\\

\begin{figure}[t]
 \includegraphics[width=0.48\textwidth]{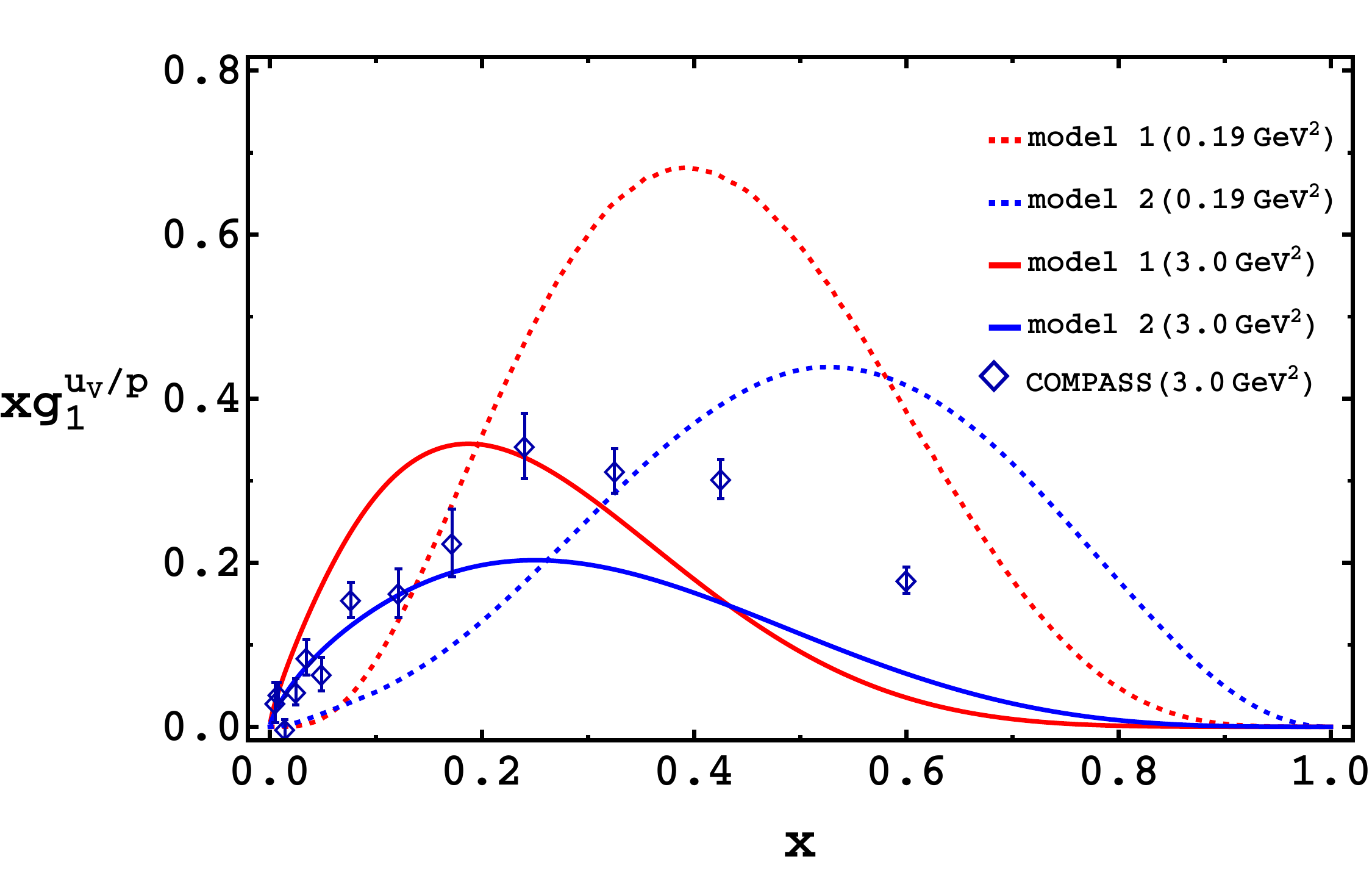}
  \hspace*{0.2cm}
  \includegraphics[width=0.48\textwidth]{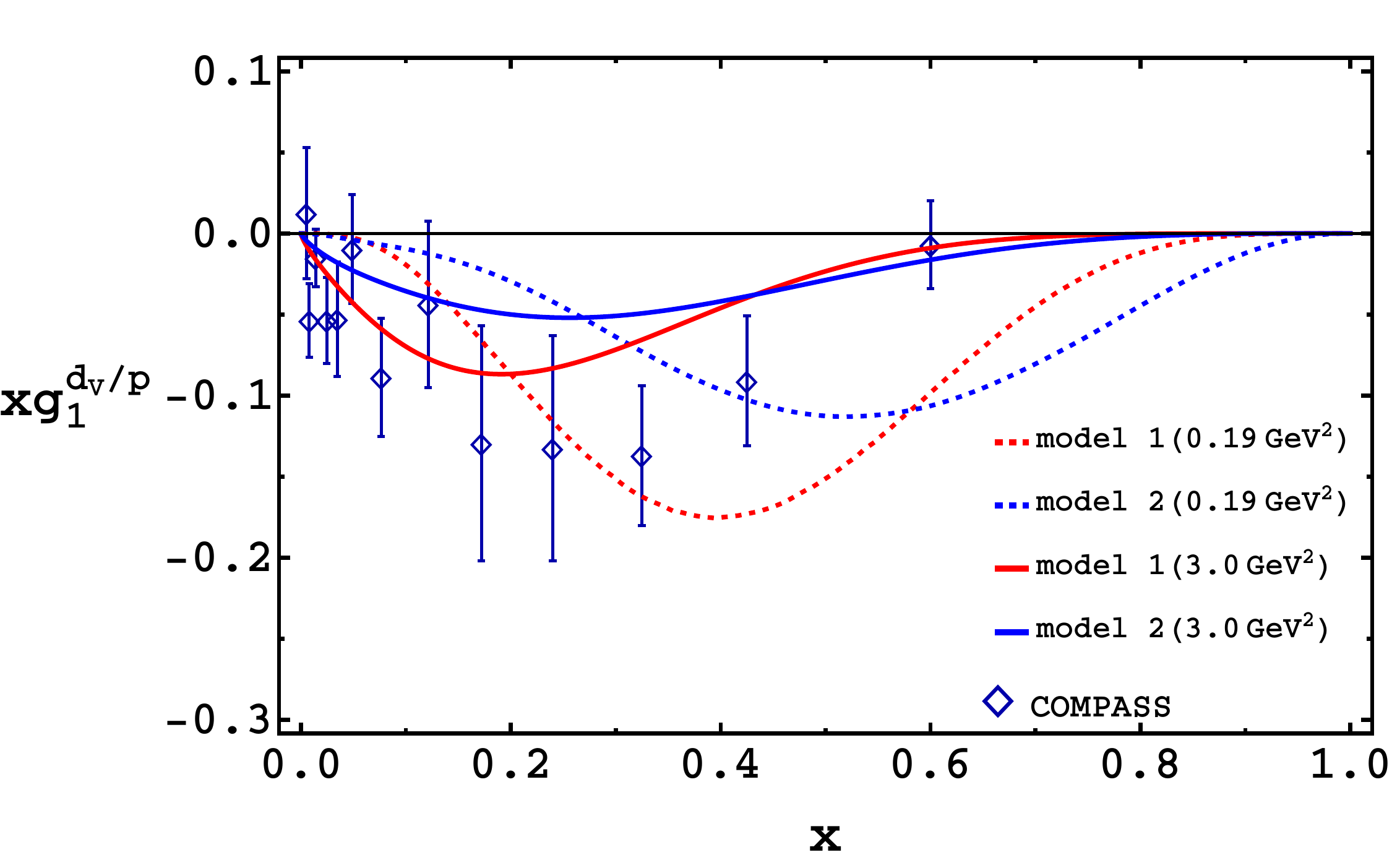}
   \includegraphics[width=0.48\textwidth]{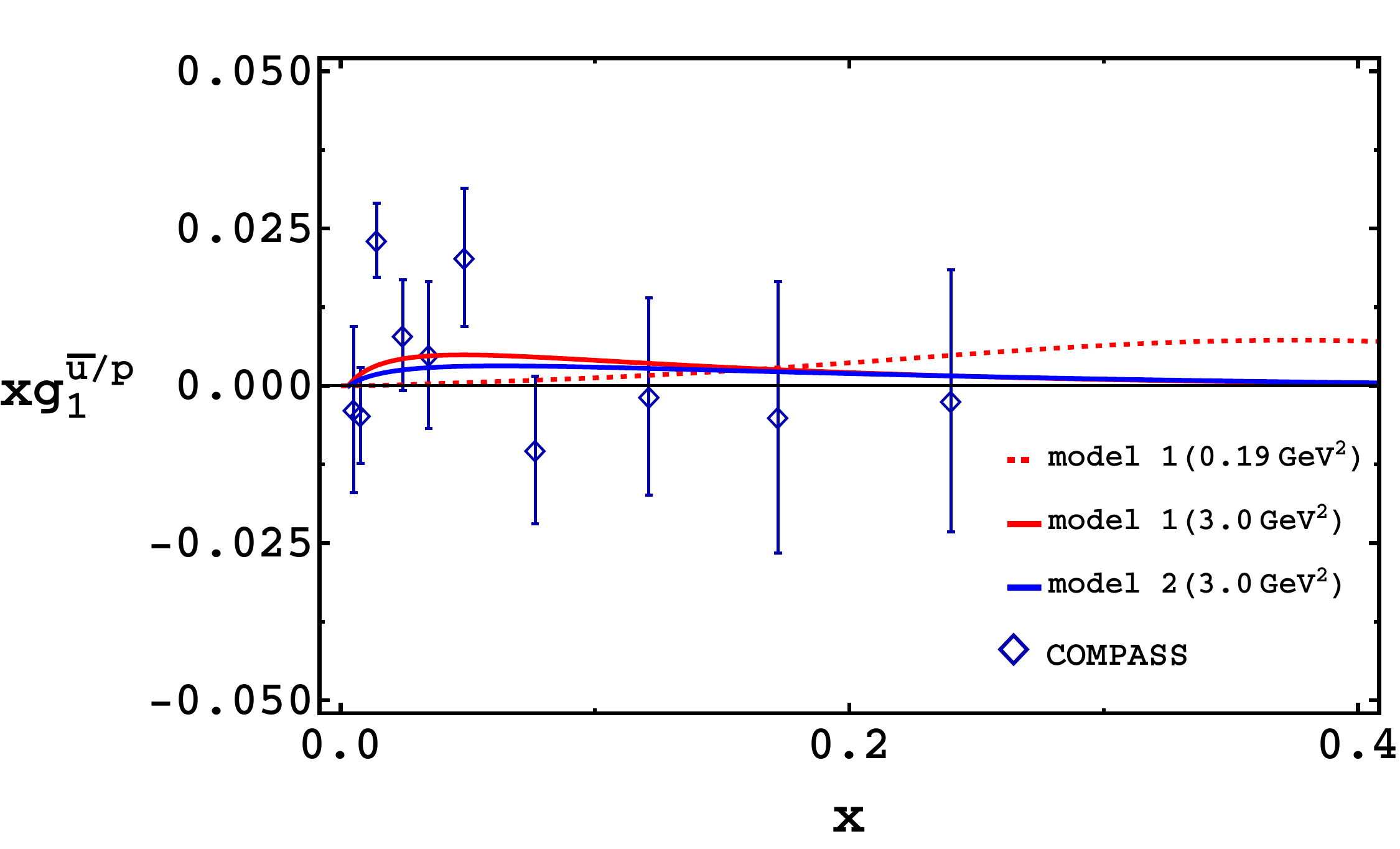}
  \hspace*{0.2cm}
  \includegraphics[width=0.48\textwidth]{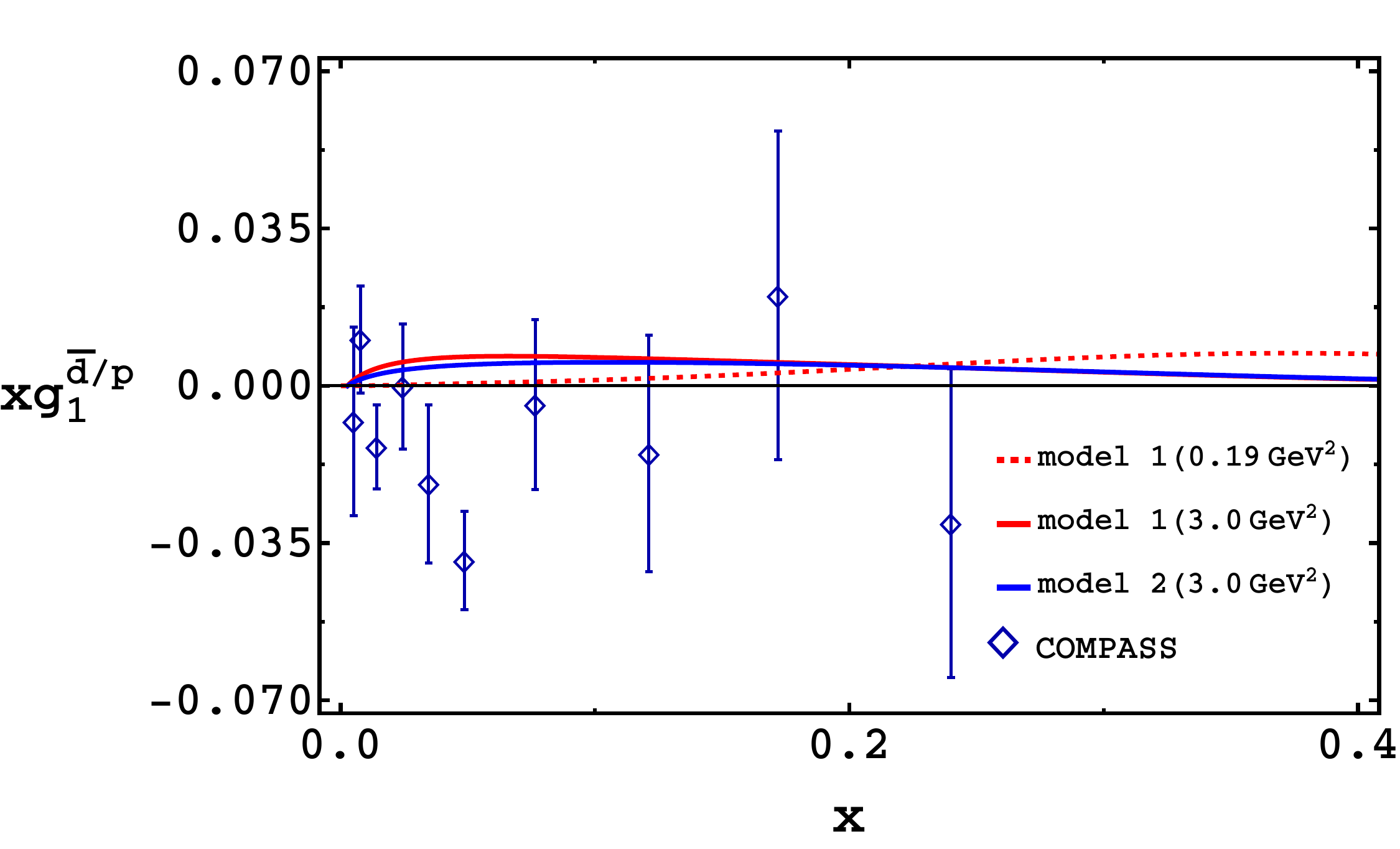}
  \caption{
  The polarized parton distribution $x g_1$ for the up (left panels) and down (right panels) quark as function of $x$.
  The upper and lower panels show the results for the valence and sea quark contributions, respectively.
The red and blue curves are obtained  within the light-front meson-cloud model from the \textit{model 1} and \textit{2} for the nucleon LFWF, respectively, at the input scale $Q^{2}_{0} $  (dotted curves) and after LO evolution to 
  $Q^{2}=3 $ GeV$^2$  (solid curves).  
  The experimental data at $Q^2= 3$ GeV$^2$ are from the COMPASS collaboration~\cite{Alekseev:2010ub}.}
   \label{fig_g1_comparsion_compass}
\end{figure}

We now turn to discuss the  distributions for longitudinally polarized proton.
In Fig.~\ref{fig_g1_comparsion_compass} we show our results for both the valence-quark and sea contributions to  the  polarized PDF \( g_1 \).
The difference between the predictions  at the input scale for the valence-quark distributions from the two models for the bare nucleon LFWFs is more pronounced than in the case of the unpolarized distributions, and persist also after LO evolution. 
On the other side, the polarized sea quark distributions are generated at the input scale only from the antiquark of the $\rho$ in the$N \rightarrow \rho N$ fluctuation, and therefore they are the same in both models at the input scale. The effect of perturbative evolution leads to a small difference in the two models.\\
The LO evolved results at $Q^2=3$ GeV$^2$ are  compared with the experimental data from COMPASS \cite{Alekseev:2010ub}.
Our predictions for the valence down-quark contribution are in fair agreement  with the experimental data, within error bars, while both models fail in reproducing the larger $x$ behaviour of the data for the up-quark contribution.
The huge error bars of the experimental data do not allow a conclusive remark about the behaviour of the sea contribution to \( g_1 \).

\begin{figure}[h!]
\vspace{1.2 truecm}
 \includegraphics[width=0.48\textwidth]{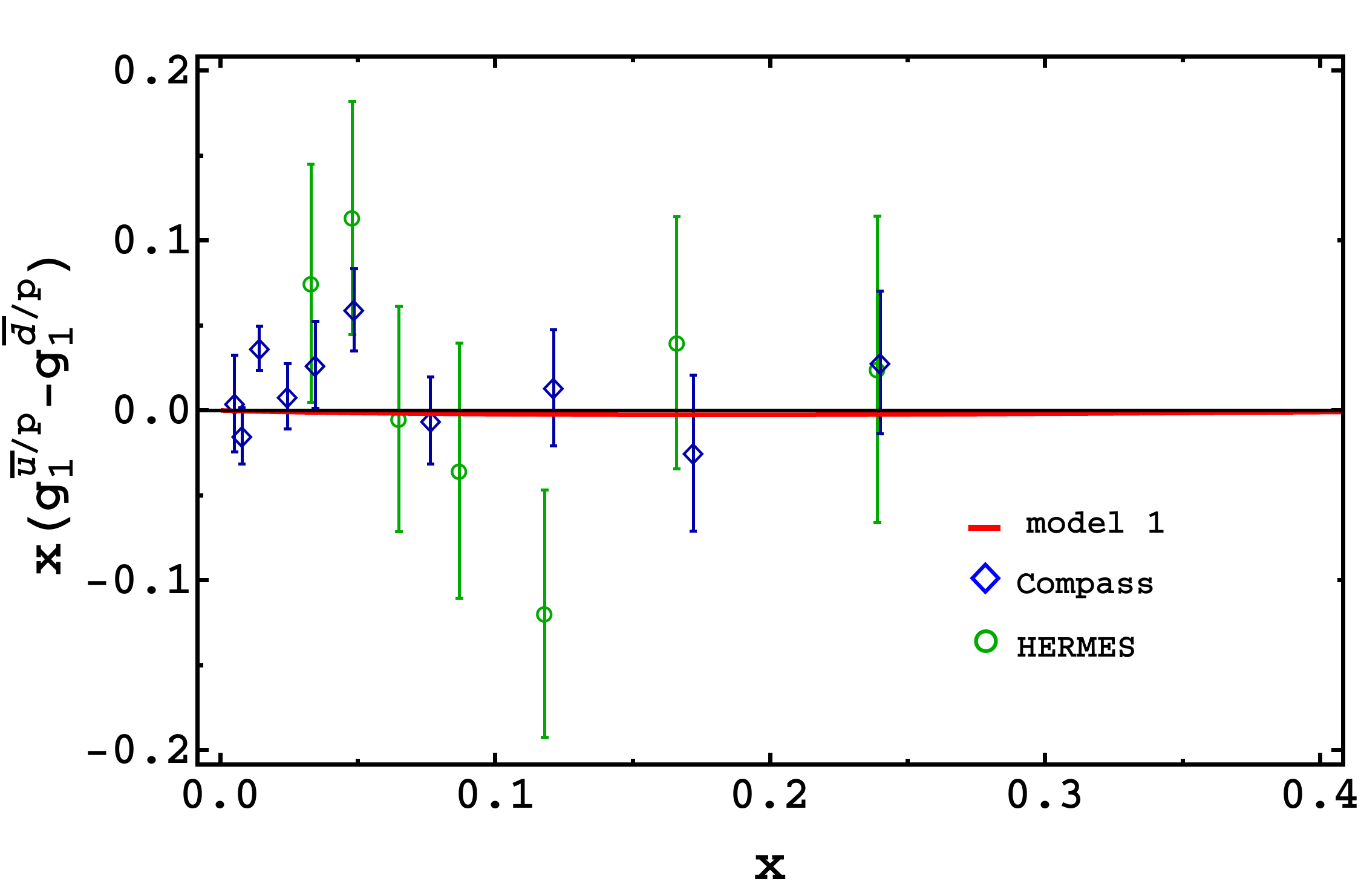}
   \caption{The flavor asymmetry of the polarized sea \(x\left( g_1^{\bar{u}/p}(x)- g_1^{\bar{d}/p}(x) \right)\) within the light-front meson-cloud model after LO evolution to $Q^2=3$ GeV$^2$ in comparison with the experimental data from COMPASS~\cite{Alekseev:2010ub} at $Q^2 =3$ GeV$^2$ and HERMES ~\cite{Airapetian:2004zf} at $Q^2=2.4$ GeV$^2$.
   The results refer to the \textit{model 1} for the nucleon LFWF, and are indistinguishable from the predictions from the \textit{model 2}. }
  \label{fig_polarised_sea_asymmetry}
\end{figure}

\noindent 
The large flavor asymmetry in the unpolarized sea naturally leads to the question whether the polarized sea is also asymmetric.
We note right away that within our model this asymmetry cannot be very large, as the $\rho$ fluctuation contributing to the polarized sea is suppressed because of the large $\rho$ mass, see Eq.~\eqref{eq_def_splitting_function}. 
Our results  for the polarized flavor asymmetry  \(x\left( g_1^{\bar{u}/p}(x)- g_1^{\bar{d}/p}(x) \right)\) are shown in Fig.~\ref{fig_polarised_sea_asymmetry}
in comparison with the experimental data from COMPASS~\cite{Alekseev:2010ub} and HERMES~\cite{Airapetian:2004zf}.
We predict a small and negative value for \(x\left( g_1^{\bar{u}/p}(x)- g_1^{\bar{d}/p}(x) \right)\) at variance with the data, that, despite the poor accuracy, seem to favour  a small positive value.
Our results are similar to the predictions within different variants of the meson-cloud model where the $\rho$ meson is responsible of the polarized sea asymmetry~\cite{Cao:2003zm,Cao:2001nu,Fries:1998at,Kumano:2001cu}.
They differ  in sign and by one of order of magnitude from the findings within quark based models, such as the chiral quark soliton model~\cite{Diakonov:1996sr,Diakonov:1997vc,Dressler:1999zg,Dressler:1999zv,Wakamatsu:2014asa} and the statistical approach~\cite{Bourrely:2001du}. 
Recently, it has been suggested that a possible way to restore this difference in meson-cloud models is to include the contributions from the interference of $\pi N$ and $\sigma N$ components in the handbag diagram of the polarized PDFs~\cite{Fries:2002um}. Such contribution has been studied at qualitative level within an effective low-energy model, by matching  the QCD operator of the polarized parton distribution with an effective operator built from the $\pi$ and $\sigma$ fields.
It would be interesting to further explore such an approach within a light-front meson-cloud model, by incorporating this effective pion and sigma components in the nucleon LFWF consistently with a light-front Fock-space expansion.\\%

\begin{figure}[t]
 \includegraphics[width=0.6\textwidth]{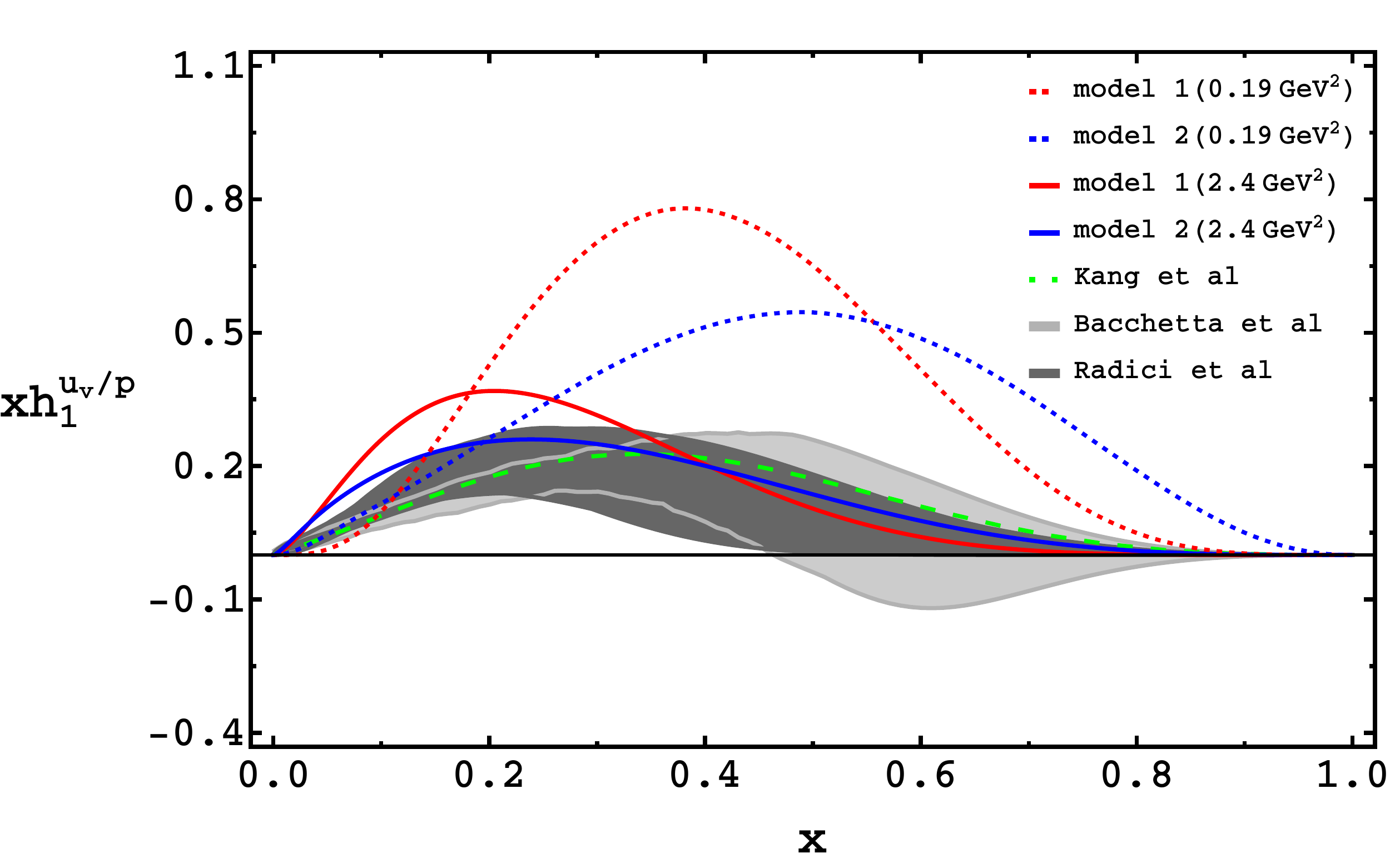}
  \includegraphics[width=0.6\textwidth]{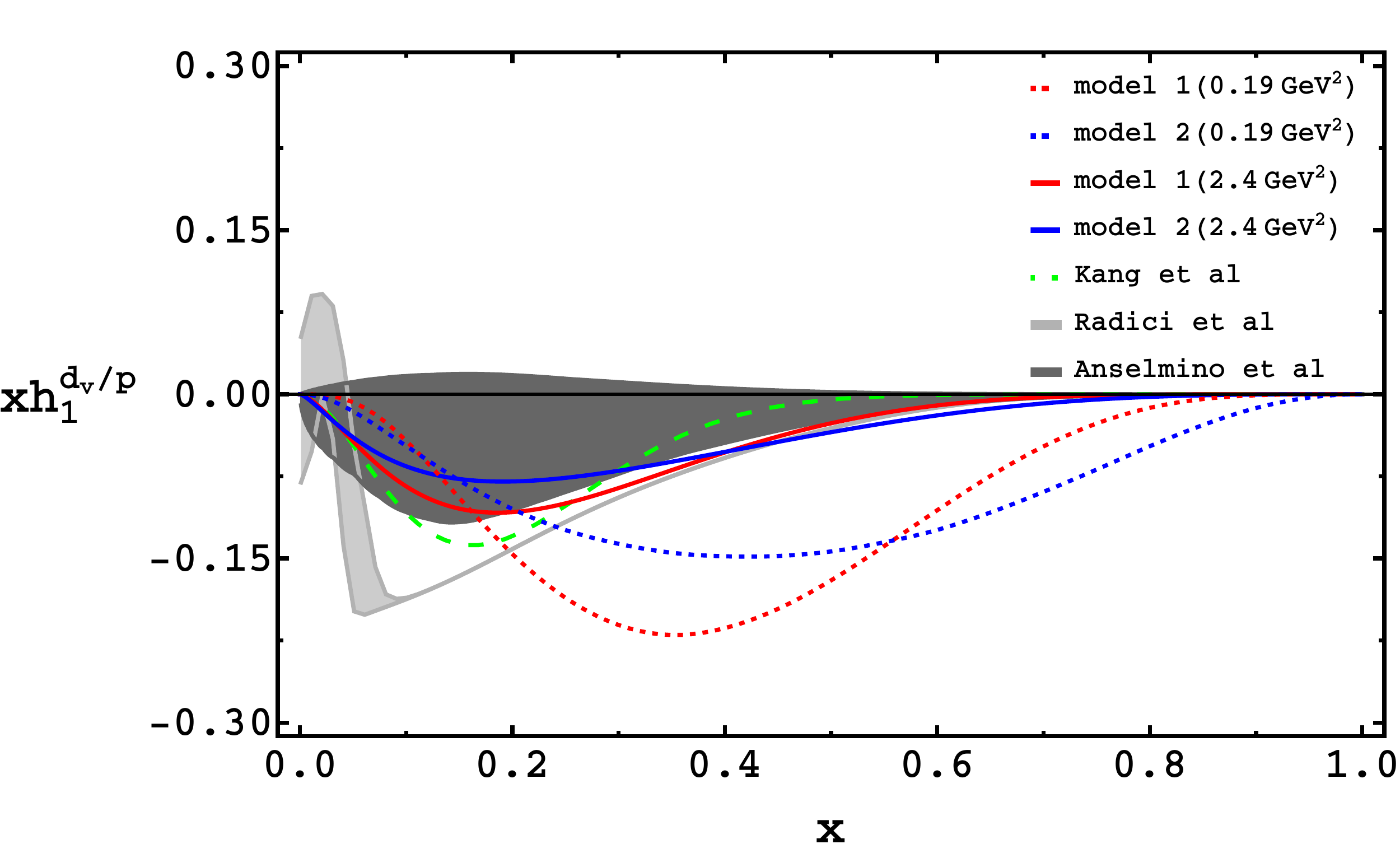}
  \caption{
  The transversity distribution $x h_1$ as function of $x$ for the valence up (upper panel) and down (lower panel) quark.
  The red and blue curves are obtained  within the light-front meson-cloud model from the \textit{model 1} and \textit{2} for the nucleon LFWF, respectively, at the input scale $Q^{2}_{0} $  (dotted curves) and after LO evolution to $Q^{2}=2.4 $ GeV$^2$  (solid curves).  
  The  phenomenological extractions at $Q^2=2.4 $ GeV$^2$ are from Ref.~\cite{Anselmino:2013vqa}  (darker bands), Ref.~\cite{Kang:2015msa} (dashed green curves), and Ref.~\cite{Radici:2015mwa} (lighter bands).
  }
  \label{fig_h1_valence}
\end{figure}

Finally, we discuss the transversity distribution.
In Fig.~\ref{fig_h1_valence} we show our model calculations for the valence transversity distribution  at the input scale \( Q_0^2 = 0.19~\mbox{GeV}^2 \) and after LO evolution to $Q^2 = 2.4~\mbox{GeV}^2$. The darker bands are the results from Ref.~\cite{Anselmino:2013vqa}, obtained by a simultaneous extraction of the transversity and Collins function from azimuthal asymmetries in SIDIS and $e^+e^-$ data, implementing evolution effects at LO in the collinear framework. The dashed green curve are the results from the extraction of Ref.~\cite{Kang:2015msa}, where evolution equations have been computed in the TMD framework at NLO.
The lighter bands refer to the extraction of Ref.~\cite{Radici:2015mwa}, where transversity has been extracted in the standard framework of collinear factorization using SIDIS with two hadrons detected in the final state, including evolution effects at LO.
 All the phenomenological extractions refer to the scale $Q^2 = 2.4~\mbox{GeV}^2$, and, despite the different frameworks of analysis and the different data sets used in the fits, give similar results for the valence up quark contribution, while show discrepancies for the down contribution, which are presumably induced by the data set of the deuteron target employed in the analysis of Ref.~\cite{Radici:2015mwa}.
Our results from the light-front meson-cloud model using different LFWFs for the bare nucleon are compatible, within error bars, with the various extractions for the up quark, and in agreement with the extractions of Refs.~\cite{Anselmino:2013vqa,Kang:2015msa} for the down quark.

 \begin{figure}[h!]
  \includegraphics[width=0.48\textwidth]{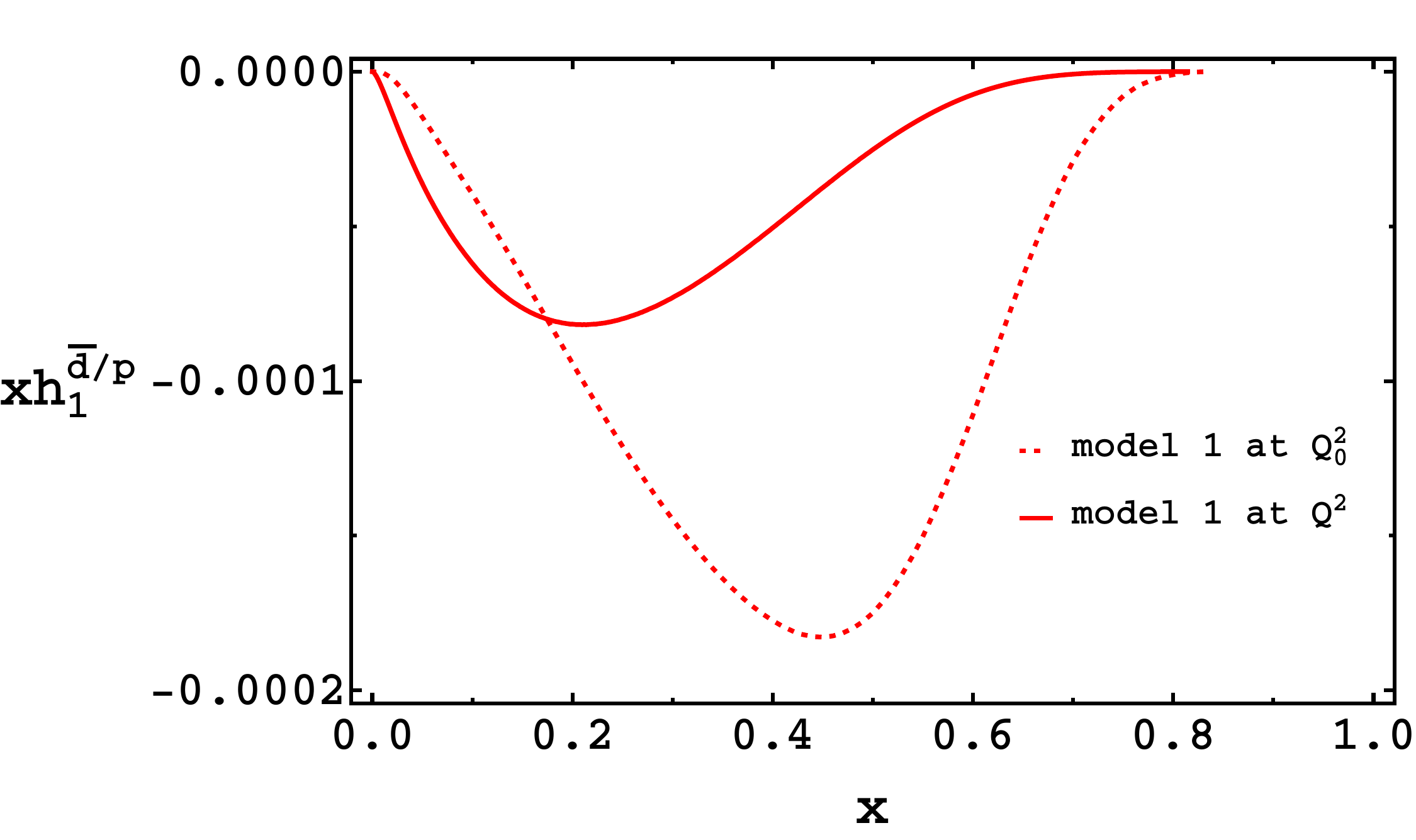}
   \hspace*{0.2cm}
   \includegraphics[width=0.48\textwidth]{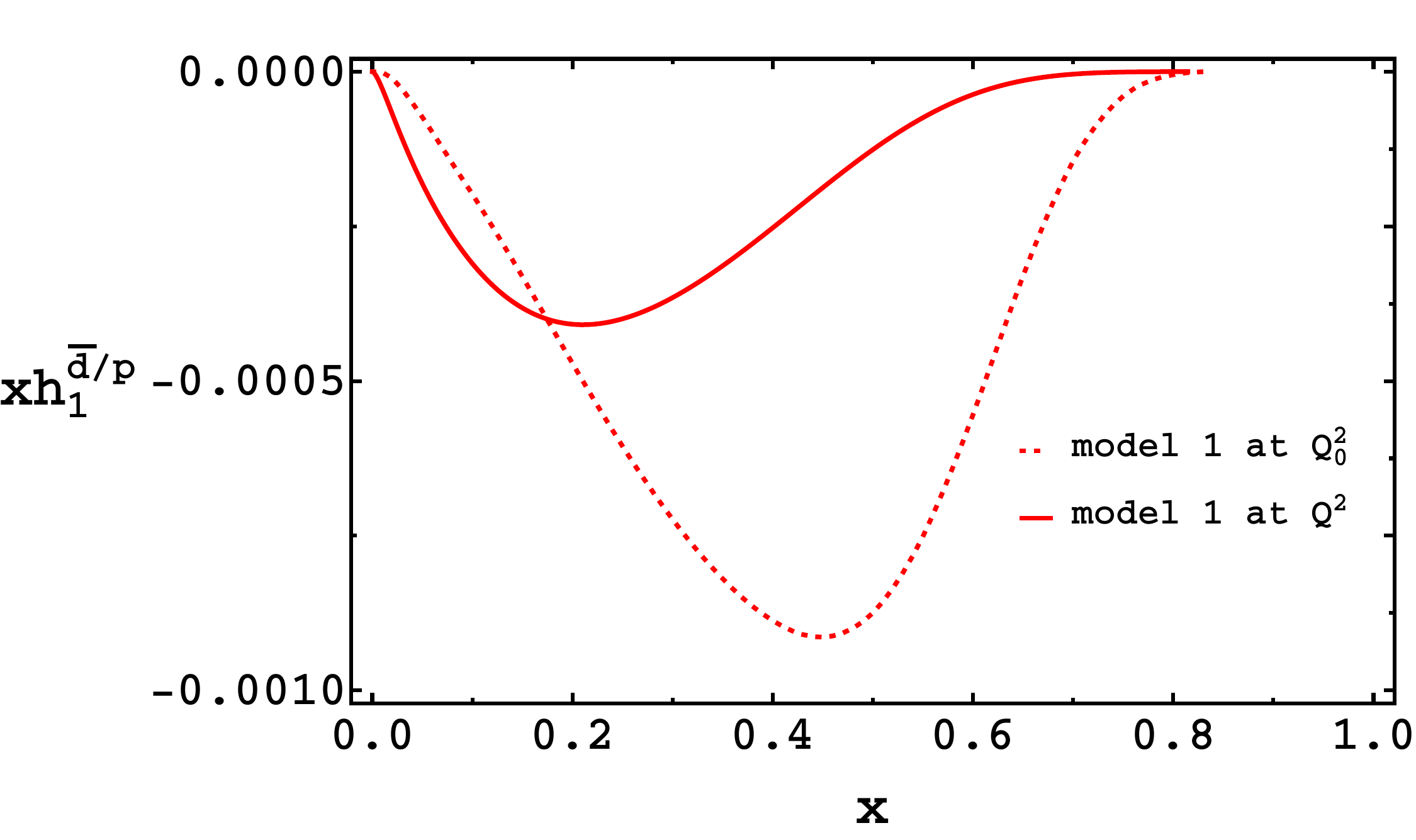}
   \caption{The transversity distribution $x h_1$ as function of $x$ for the up (left panel) and down (right panel) antiquark. The dotted curves show the results at the input scale $Q^{2}_{0} $, and the solid curves are the results after LO evolution to $Q^2=2.4 $ GeV$^2$.
  }
   \label{fig_h1_sea}
 \end{figure}
\begin{figure}[h]
\centering
 \includegraphics[width=0.49\textwidth]{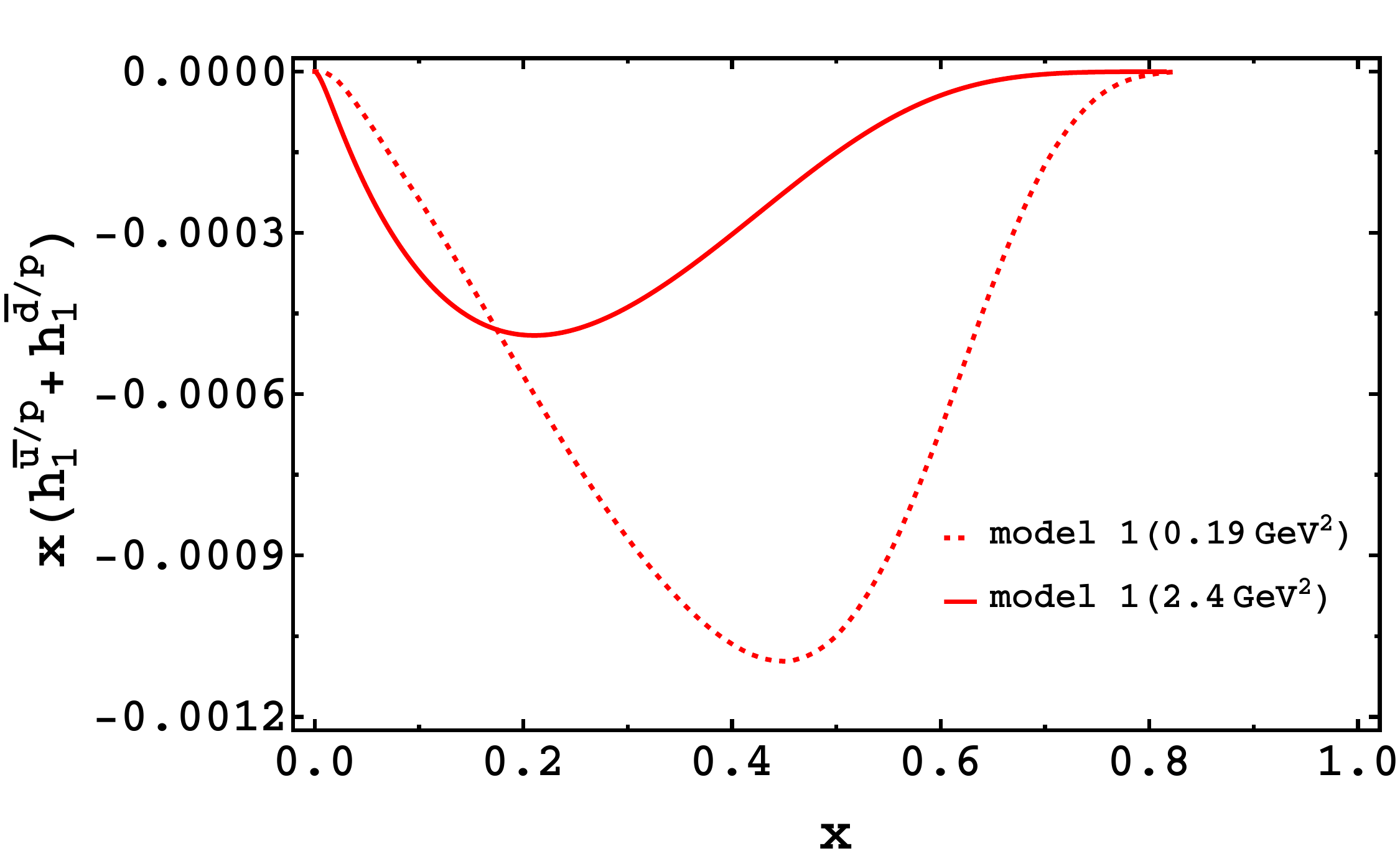}
%   \hspace*{0.2cm}
  \includegraphics[width=0.48\textwidth]{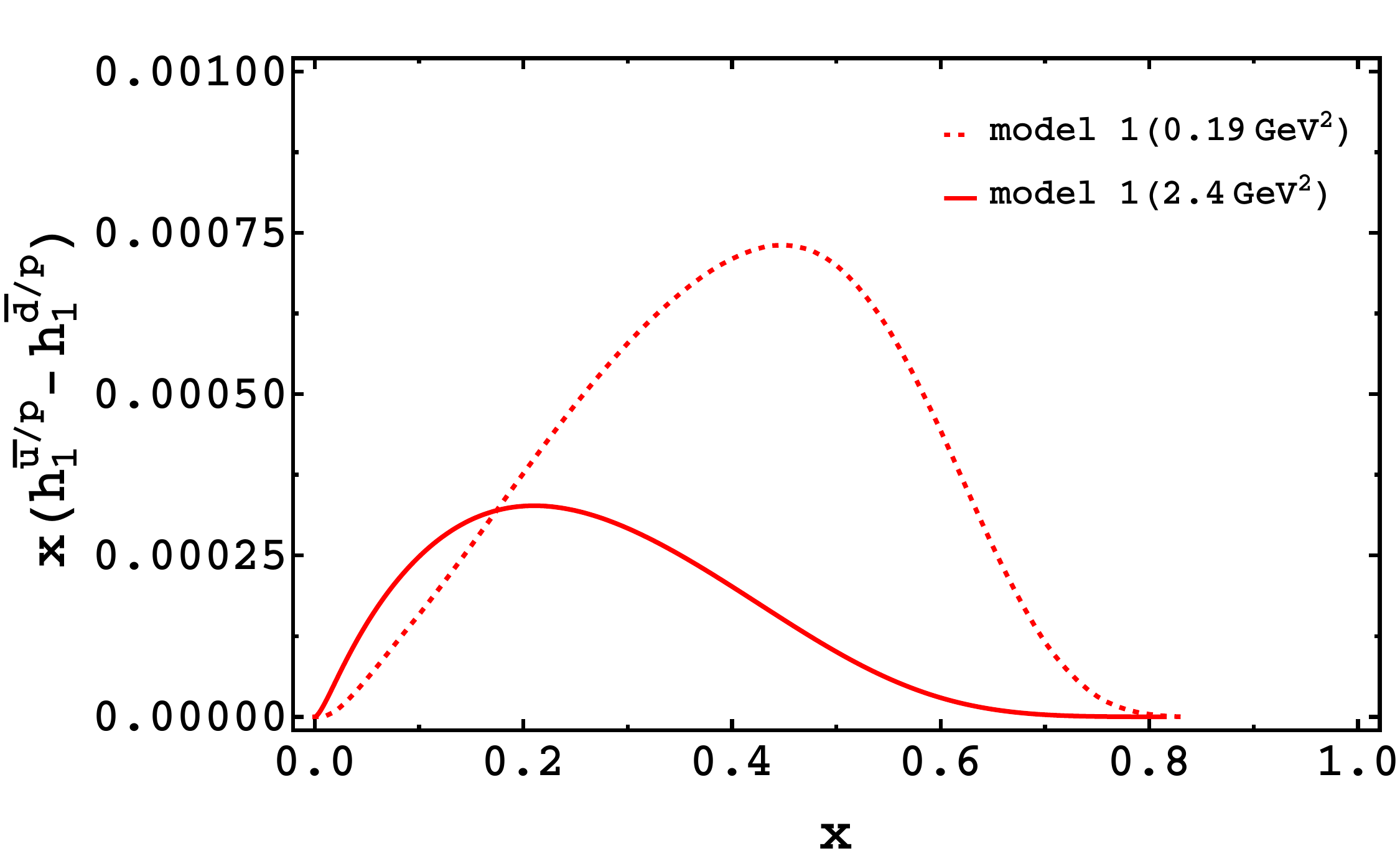}
   \hspace*{0.4cm}
   \includegraphics[width=0.47\textwidth]{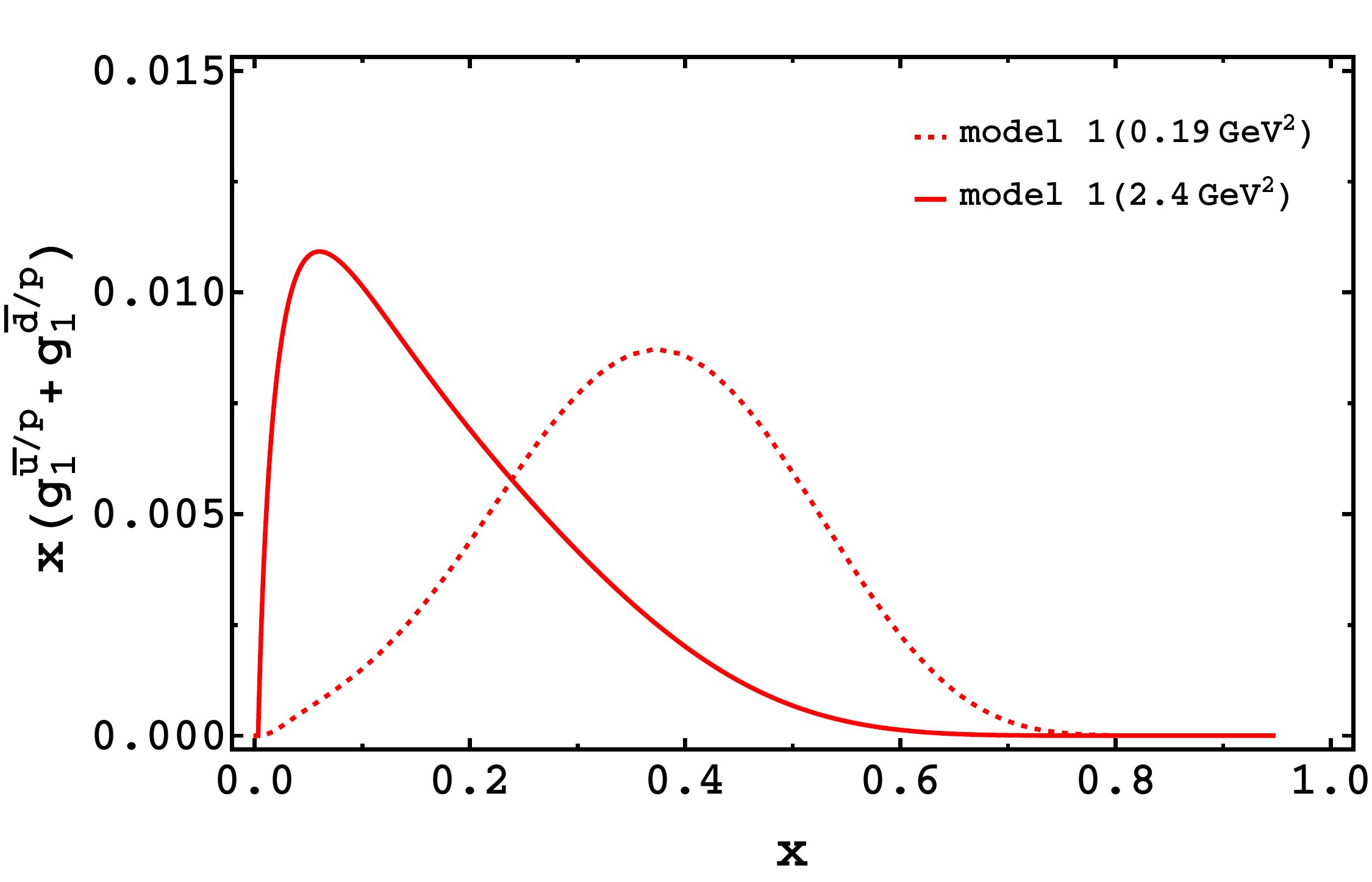}
  \hspace*{0.1cm}
  \includegraphics[width=0.47\textwidth]{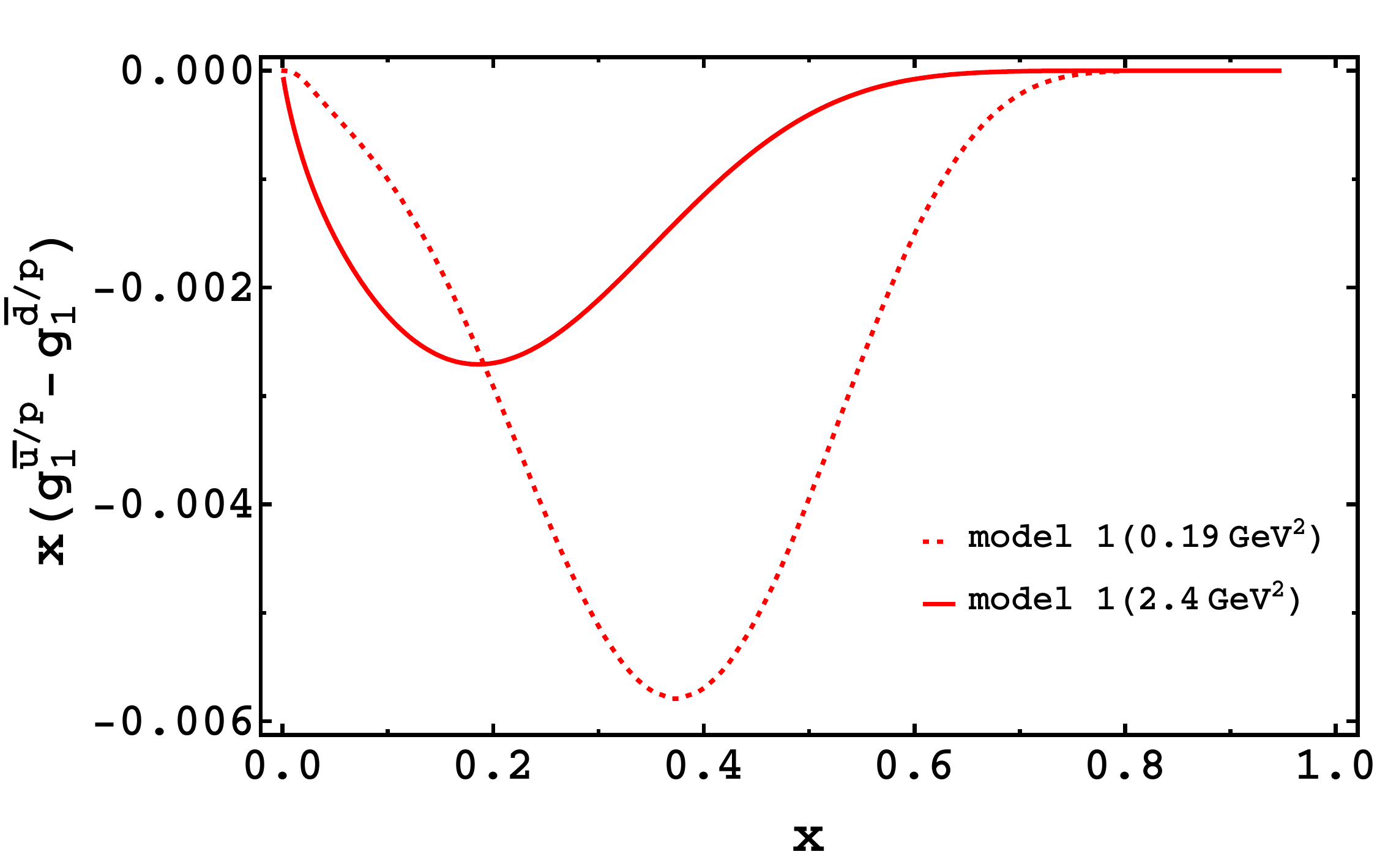}
 \caption{The transversity (upper panels) and longitudinally polarized (lower panels) antiquark distributions as function of $x$ from the light-front meson-cloud model with \textit{model 1} for the nucleon LFWF.
  The left panels show the isoscalar contributions for the flavor combination $\bar u+\bar d$, and the right panels are the isovector contributions for  the flavor combination $\bar u- \bar d$.
  The dotted curves refer to the model results at the input scale $Q^{2}_{0}$, while the solid curve are the model results after LO evolution to $Q^2=2.4$ GeV$^2$.}
  \label{fig_meson_cloud_iso_vector}
\end{figure}
\noindent The results for the sea quark contribution to the transversity are  shown in Fig.~\ref{fig_h1_sea}.
The transversely polarized sea is generated only by the $\rho$ fluctuations, and evolves independently of the quark contribution, thanks to the chiral-odd nature of the transversity. Therefore the sea distributions are independent on the model for the bare nucleon LFWF both at the input scale and after LO evolution. The results are very small, even one order of magnitude smaller than in the case of the longitudinally polarized distributions.
Furthermore, we find negative results for both the up and down antiquark, with a larger contribution in absolute value for the down antiquark.
These predictions are the first calculations for the antiquark transversity distributions within a meson-cloud approach.  We can compare our results with different approaches in literature, such as the chiral quark soliton model~\cite{Schweitzer:2001sr} and a chiral chromodielectric model~\cite{Barone:1996un,Barone:2003jp}.
In the chromodieletric model, the transversity antiquark distribution is also quite small, although it differs both in sign and in the relative magnitude of the anti-up and anti-down contributions with respect to our predictions.
On the other side, the chiral quark soliton model predicts much larger contributions.
According to the expectation from  the large $N_c$ limit,  the chiral quark soliton model satisfies the following inequalities~\cite{ Pobylitsa:1996rs,Pobylitsa:1998tk,Pobylitsa:2000tt}

\bea
|g_{1}^{\bar u}-g_{1}^{\bar d}|>|g_{1}^{\bar u}+g_{1}^{\bar d}|,\nonumber \\
|h_{1}^{\bar u}-h_{1}^{\bar d}|>|h_{1}^{\bar u}+h_{1}^{\bar d}|.
\label{large-nc}
\eea
Analogous relations hold for the quark distributions.
In the case of the antiquark distributions, the hierarchy in Eqs.~\eqref{large-nc}  is not supported from  both the chromodielectric model and the light-front meson-cloud model (see Fig.~\ref{fig_meson_cloud_iso_vector}), while it holds for the quark contributions.\\
\begin{figure}[!t]
 \includegraphics[width=0.7\textwidth]{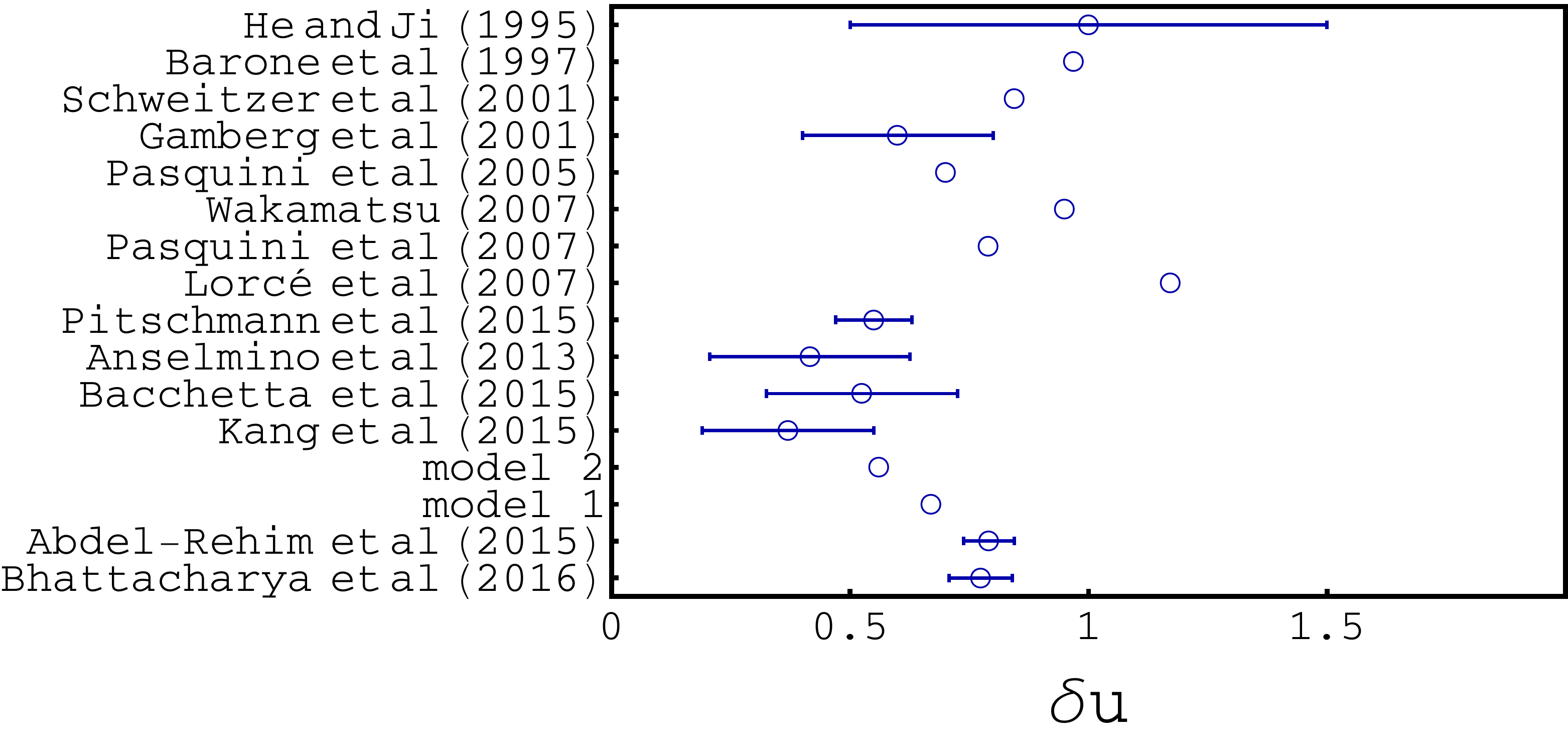}\\
 \vspace{1 truecm}
  \includegraphics[width=0.7\textwidth]{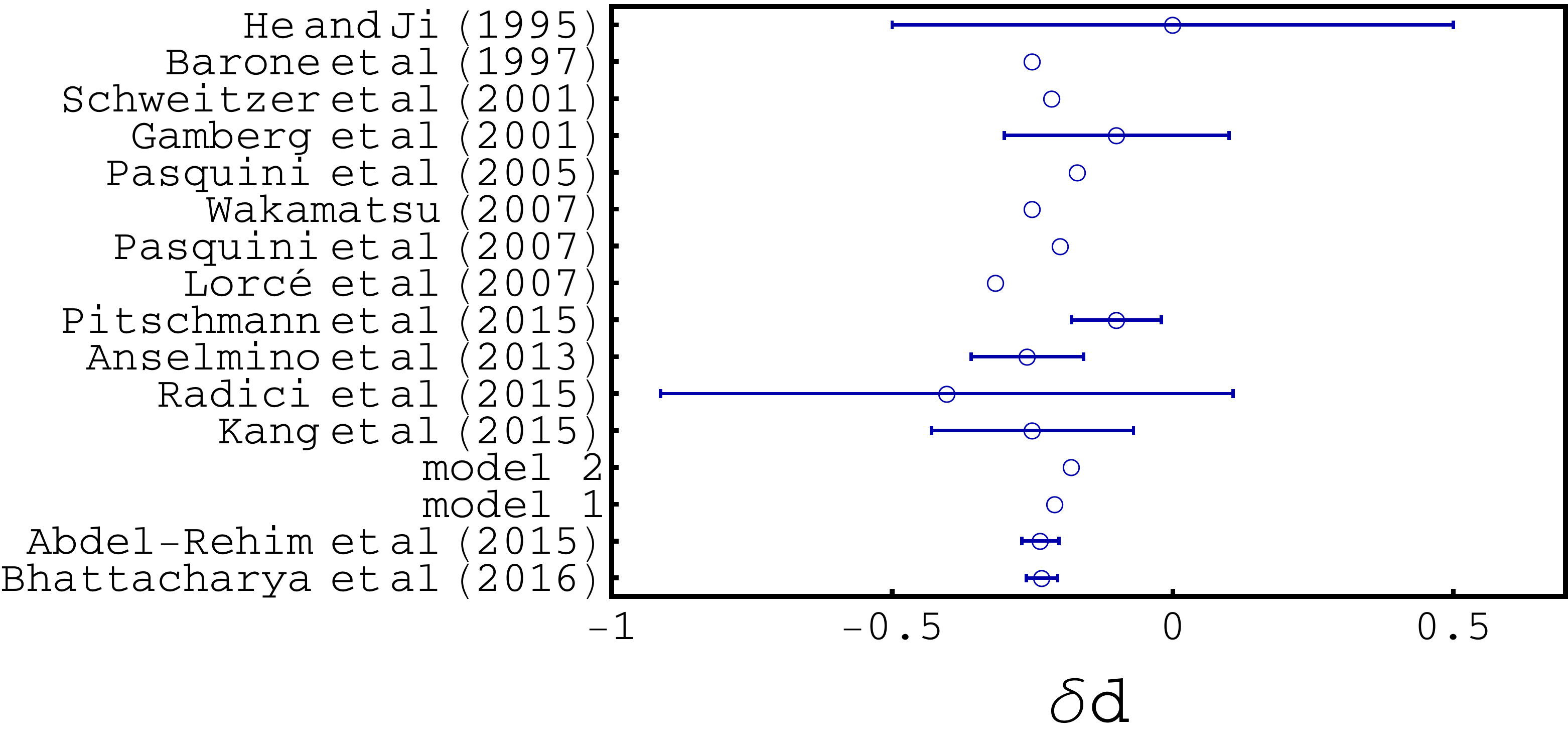}
   \caption{Comparison of our model calculation of the tensor charges of up (upper panel) and down quark (lower panel) after LO evolution to $Q^2=10$ GeV$^2$ with the results from 
   different models and phenomenological extractions:
    He and Ji \cite{He:1994gz} (\(  Q^2  \sim 1 ~\mbox{GeV}^2\)), Barone et al.~\cite{Barone:1996un} ($Q^2=25$ GeV$^2$), Gamberg et al. \cite{Gamberg:2001qc} (\(  Q^2 \sim1~\mbox{GeV}^2\)), Pasquini et al. 2005 \cite{Pasquini:2005dk} (\(  Q^2 = 10~\mbox{GeV}^2\)), Wakamatsu \cite{Wakamatsu:2007nc} (\(  Q^2 = 2.4~\mbox{GeV}^2 \)), Pasquini et al. 2007 \cite{Pasquini:2006iv,Pincetti:2006hc} (\(  Q^2 = 10~\mbox{GeV}^2\)), Lorc\'e~\cite{Lorce:2007fa} ($Q^2=0.36$ GeV$^2$, Pitschmann et al.\cite{Pitschmann:2014jxa} (\( Q^2 = 4~\mbox{GeV}^2\)),  Anselmino et al. \cite{Anselmino:2013vqa} (\( Q^2 = 0.8~\mbox{GeV}^2\)), Radici et al. \cite{Radici:2015mwa} (\( Q^2 =10~\mbox{GeV}^2\)), Kang et al. \cite{Kang:2015msa} (\( Q^2 = 10~\mbox{GeV}^2\)), our results within \textit{model 1} and \textit{2} for the bare nucleon LFWF ($Q^2=10$ GeV$^2$), Abdel-Rehim et al.~\cite{Abdel-Rehim:2015owa} ($Q^2=4$ GeV$^2$),     Bhattacharya et al.~\cite{Bhattacharya:2016zcn} ($Q^2=4$ GeV$^2$).  
   } 
  \label{fig_tensor_charge}
\end{figure}

\noindent The first moment of the flavor non-singlet combination of the transversity  gives the quark tensor charge:
\bea
\delta q=\int_{0}^{1} {\rm d}x\, (h_{1}^{q}(x)-h_{1}^{\bar q}(x)).
\eea 
In Fig.~\ref{fig_tensor_charge} we collect the results for the up and down quark tensor charges  from the light-front meson-cloud model along with the estimates  from other theoretical approaches and   data analysis.
The results in Ref.~\cite{Pasquini:2005dk} and Refs.~\cite{Pasquini:2006iv,Pincetti:2006hc} 
have been obtained by taking into account only the three-quark component of the nucleon state, using \textit{model 1} and  \textit{model 2} for the bare nucleon LFWF, respectively.
Therefore, comparing these results with the present calculations in the meson-cloud model, we can estimate the  effects of including the five-parton component in the nucleon LFWF. 
This comparison in Fig.~\ref{fig_tensor_charge} is performed after evolution of the results to the scale $Q^2=10$ GeV$^2$, which coincide with the scale of the phenomenological extractions~\cite{Anselmino:2013vqa,Radici:2015mwa,Kang:2015msa}.
We notice that the inclusion of the meson-cloud contribution give small corrections, improving the agreement of our results with the data analysis, especially for the up quark.
The results in other theoretical frameworks have been reported at different scales, as given in the original works. They correspond to: $Q^2=1$ GeV$^2$ in the QCD sum rule approach~\cite{He:1994gz} and the axial-vector dominance model~\cite{Gamberg:2001qc}; $Q^2=2.4$ GeV$^2$ in the chiral-quark soliton model  of Ref.~\cite{Wakamatsu:2007nc}; $Q^2=4$ GeV$^2$ in the  Dyson-Schwinger model~\cite{Pitschmann:2014jxa} and lattice QCD calculations~\cite{Abdel-Rehim:2015owa,Bhattacharya:2016zcn}; $Q^2=0.36$ GeV$^2$ in the chiral quark soliton model calculation of Ref.~\cite{Schweitzer:2001sr} and the light-front chiral quark soliton model, truncated to the 5-parton component in the Fock space, of Ref.~\cite{Lorce:2007fa}; $Q^2=25$ GeV$^2$ in the chromodielectric model of Ref.~\cite{Barone:1996un}.
However, the dependence on the scale of the tensor charge is quite weak, and in general all the model calculations are consistent with the data analysis for the down-quark tensor charge, while  distinctions among the various models appear for the up-quark contribution.

\section{Conclusions}
\label{sect:6}
The convolution model for the physical nucleon, where the bare nucleon is dressed by its
virtual meson cloud, has seen a wealth of applications to describe the non-perturbative origin of the sea-quark structure.
In this paper this approach has been revisited and applied to the leading-twist collinear parton distribution functions (PDFs) within a light-front formalism, in particular discussing the formalism necessary for the calculation of thevalence- and sea quark transversity distribution.
The  dressing of the physical nucleon is obtained through fluctuations of the bare nucleon into baryon and meson states, which are calculated in the one-meson approximation, using light-front time ordered perturbation theory.
Furthermore, the bare baryon and meson states are described in terms of light-front wave functions (LFWFs), taking into account the corresponding valence parton configurations. 
In the explicit calculation, we consider baryon-meson fluctuations with the baryon being a nucleon and the meson being a pion either a rho.
Within this model, the sea contribution can be generated only from the antiquark constituent of the mesons.
In particular, both the pion and rho participate to the unpolarized sea distributions, while in the case of longitudinally and transversely polarized sea, only the vector meson rho contributes.
As the probability amplitude for the nucleon to fluctuate  into a baryon-meson state depends on the inverse of the squared invariant mass of the baryon-meson state,  the contribution of the rho is suppressed with respect to the pion.
Accordingly, the polarized  sea quark distributions are much smaller than the unpolarized sea distributions and the sea-quark contribution to the transversity is even more suppressed.
The 3-quark component of the nucleon state has been described using two different models for the LFWF.
This component gives the main contribution to the valence part of the parton distributions and produces quite different predictions at the hadronic scale within the two models.
\\
QCD evolution effects have been taken into account applying standard DGLAP equations at leading order (LO).
In the case of the unpolarized and transversity distributions, the differences between the two models for the valence contributions become much smaller after evolution to higher scales and the results in both models are well compatible with the available parameterizations and phenomenological extractions.
In the case of the helicity distribution, the differences between the two models remain more pronounced also after LO evolution, especially for the valence up contribution where the agreement with the available experimentally data is not very satisfactory.

 Starting with the same input for the sea quark contribution, the different evolution equations for the non-singlet and singlet combinations of the unpolarized and helicity distributions generate different results for the sea distributions at higher scales, when using the two different models for the bare nucleon LFWF.
Vice versa, in the case of the transversity, the non-singlet and singlet evolution equations are equal, and therefore the distribution of the sea-quark transversity is independent on the model for the bare nucleon LWFW both at the input and higher scales.
The results for the unpolarized sea-quark distributions confirm the findings of previous calculations within different variants of the meson-cloud model, where the excess of the $\bar d$ over $\bar u$ has a pure non-perturbative origin at the input scale of the model and is able to to explain the main features of the observed sea-quark flavor asymmetry after QCD evolution to the relevant experimental scales. 
The situation for the ratio of the $\bar d$ over the $\bar u$ unpolarized parton distributions is less clear. In this case, the experimental data become less precise at larger values of $x$.
However, they seem compatible with a rapid decrease of this ratio towards and below unity at larger values of $x$, at variance with the predictions within our light-front  meson cloud model and the findings of previous meson-cloud models and other non-perturbative models, %such as chiral perturbation theory and instanton models, 
as well as the PDF fit provided by a statistical  model.
New results  from the recent Drell-Yan measurements of the Fermilab E906/SeaQuest experiment are expected soon, and hold the promise to reduce the experimental uncertainties of the data at higher values of $x$, providing  a better understanding of the behavior of the ratio in this region.
We also confirm the findings within different variants of the meson-cloud approach which 
predict a very small flavor asymmetry of the longitudinally polarized  sea-quark distributions, with an excess of the $\bar u$ over the $\bar d$ contribution. These results are at variance with the experimental data, that, despite the poor accuracy, seem to favor the opposite trend for the sign of the polarized flavor asymmetry.

The results for the sea-quark transversity distribution within a meson-cloud approach are discussed here for the first time. The small sea-quark contribution at the input scale is further suppressed after evolution to larger scales.
The sign for the flavor asymmetry of the sea-quark transversity distributions is opposite to the predictions from different approaches such as the chromodielectric  model and the quark soliton model.
On the other side, the relative order of magnitude for the absolute values of the helicity and transversity flavor asymmetry is the same in our model and the chromodielectric model, but it is not consistent with the large $N_c$ expectations of the quark soliton model.

Finally, we discussed the results for the tensor charge of the nucleon. Although the effects of introducing the meson-cloud contribution to the three-quark component of the nucleon state are small, they improve the agreement of our results with recent phenomenological extractions. 
This is more evident in the case of the up quark contribution, where in general the differences among different quark model calculations are more pronounced.

More elaborated meson-cloud models for the sea-quark  transversity distributions could be discussed introducing the effects of higher-mass baryon-meson fluctuations beyond the nucleon-rho contribution. However, such contributions with larger invariant mass are expected to be further suppressed with respect to the already small contribution from the nucleon-rho component, and would not change the overall findings of our model calculation.

\section*{Acknowledgements}
We thank A. Bacchetta, C. Pisano, M. Radici and W. Schweiger for helpful discussions.  This work was partially supported by the European Research Council (ERC) under the European Union's Horizon 2020 research and innovation programme (grant agreement No. 647981, 3DSPIN).
S.K. is supported by the Fonds zur F\"orderung der wissenschaftlichen Forschung in \"Oster\-reich via FWF DK W1203-N16.

\appendix

\section{Definition of parton distribution functions}
\label{appendix:a}

We introduce the following definition of the quark-quark correlator for a hadron target $H$  \begin{equation}
\Phi_{ab}(x,S)=\int\frac{\ud\xi^-}{2\pi}\,e^{ik^+\xi^-}\langle P,S; H|\overline\psi_b(0)\psi_a(\xi)|P,S;H\rangle\big|_{\xi^+=\mathbf{\xi}_\perp=0},
\label{correlator}
\end{equation}
where $k^+=xP^+$, and $\psi$ is the quark field operator with $a,b$ indices in the Dirac space. The target state is characterized by its four-momentum $P$ and covariant spin four-vector $S$ satisfying $P^2=M^2$, $S^2=-1$ and $P\cdot S=0$. We choose a reference frame where the hadron momentum has no transverse components $P=\left[P^+,\tfrac{M^2}{2P^+},\uzero_\perp\right]$, and so $S=\left[S_z\,\tfrac{P^+}{M},-S_z\,\tfrac{M}{2P^+},\uS_\perp\right]$ with $\uS^2=1$. From now on, we replace the dependence on the covariant spin four-vector $S$ by the dependence on the unit three-vector $\uS=\left(\uS_\perp,S_z\right)$.
The parton distribution functions can be obtained by performing the trace of the correlator \eqref{correlator} with suitable Dirac matrices.
Using the abbreviation $\Phi^{[\Gamma]}\equiv {\rm Tr} [\Phi\Gamma]/2$, we have
\begin{eqnarray}
\Phi^{[\gamma^+]}(x, \mathbf S)&=&f_1,\\
\Phi^{[\gamma^+\gamma_5]}(x, \mathbf S)&=&S_z g_1,\\
\Phi^{[i\sigma^{j+}\gamma_5]}(x, \mathbf S)&=&S^j_\perp h_1.
\end{eqnarray}

It is convenient to represent the correlator in terms of light-front helicity amplitudes, which treat in a symmetric way both quark and target polarization:

 \begin{equation}
\label{eq_def_hel_amps1}
 A_{\Lambda^\prime\lambda^\prime ,\Lambda\lambda}^{q/H} = \int \frac{\mathrm{d} z^-}{2\pi} e^{i \bar{x} \, \bar{p}^+ \, z^-} \braket{p,\Lambda^\prime;N|~ \mathscr{O}_{\lambda^\prime\lambda}^q  |p,\Lambda;N}|_{z^+=\mathbf z_\perp=0},
 \end{equation}

 where the quark field operator are defined by
\begin{align}
 \begin{aligned}
 \mathscr{O}_{++}^q &= \frac{1}{4}\bar{\psi}^q \left(-\frac{z}{2} \right)  \gamma^+ \left( 1 + \gamma_5 \right) \psi^q \left( \frac{z}{2} \right), \\ 
 \mathscr{O}_{--}^q &= \frac{1}{4} \bar{\psi}^q \left(-\frac{z}{2} \right) \gamma^+ \left( 1 - \gamma_5 \right) \psi^q \left( \frac{z}{2} \right),\\
 \mathscr{O}_{-+}^q &= -\frac{i}{4}\bar{\psi}^q\left(-\frac{z}{2} \right) \sigma^{+1} \left( 1 + \gamma_5 \right) \psi^q \left( \frac{z}{2}\right) \\ 
 &=-\frac{i}{4}\bar{\psi}^q\left(-\frac{z}{2} \right) \left(\sigma^{+1} -i \sigma^{+2}\right) \psi^q \left( \frac{z}{2}\right), \\ 
 \mathscr{O}_{+-}^q &= \frac{i}{4}\bar{\psi}^q\left(-\frac{z}{2} \right) \sigma^{+1} \left( 1 - \gamma_5 \right) \psi^q \left( \frac{z}{2}\right), \\ 
 &= \frac{i}{4}\bar{\psi}^q\left(-\frac{z}{2} \right) \left(\sigma^{+1}+i\sigma^{+2} \right) \psi^q \left( \frac{z}{2}\right).
 \end{aligned}
\end{align}

The decomposition of the helicity amplitudes in terms of PDFs can be obtained by decomposing the target states $|P,\pm \mathbf S; H\rangle$ with light-cone polarization parallel or opposite to
the generic direction $\mathbf S = (\sin \theta_S\cos\phi_S, \sin \theta_S \sin\phi_S, \cos\theta_S)$ in terms of the target light-cone
helicity states $|P,\Lambda; H\rangle$, 
\begin{equation}
\begin{pmatrix}|P,+\uS\rangle,&|P,-\uS\rangle\end{pmatrix}=\begin{pmatrix}|P,+\rangle,&|P,-\rangle\end{pmatrix}u(\theta_S,\phi_S),
\end{equation}
where the $SU(2)$ rotation matrix $u(\theta_S,\phi_S)$ is given by
\begin{equation}\label{su2rot}
u(\theta_S,\phi_S)=\begin{pmatrix}\cos\tfrac{\theta_S}{2}\,e^{-i\phi_S/2}&-\sin\tfrac{\theta_S}{2}\,e^{-i\phi_S/2}\\\sin\tfrac{\theta_S}{2}\,e^{i\phi_S/2}&\cos\tfrac{\theta_S}{2}\,e^{i\phi_S/2}\end{pmatrix}.
\end{equation}

For a spin $1/2$ target like the proton one has:
\begin{equation}\label{amplPDFs1/2}
A^{q/p}_{\Lambda^\prime\lambda^\prime,\Lambda\lambda}=\begin{pmatrix}
\tfrac{1}{2}\left(f^{q/p}_1+g^{q/p}_{1}\right)& 0&0 &h^{q/p}_1\\
0&\tfrac{1}{2}\left(f^{q/p}_1-g^{q/p}_{1}\right)&0&0\\
0&0&\tfrac{1}{2}\left(f^{q/p}_1-g^{q/p}_{1}\right)&0\\
h^{q/p}_1&0&0&\tfrac{1}{2}\left(f^{q/p}_1+g^{q/p}_{1}\right)
\end{pmatrix},
\end{equation}
where the rows entries are $(\Lambda^\prime\lambda^\prime)=(++),(+-),(-+),(--)$ and the columns entries are likewise $(\Lambda\lambda)=(++),(+-),(-+),(--)$

For a spin $1$ target like the $\rho$ one has \cite{Bacchetta:2001rb}\footnote{Note that 
the definition of the helicity amplitudes in Ref.~\cite{Bacchetta:2001rb} differs from our definition in Eq.~(\ref{amplPDFs1}) by a factor of two.}:

\begin{equation}
\label{amplPDFs1}
A^{q/\rho}_{\Lambda^\prime \lambda^\prime,\Lambda \lambda}=
\frac{1}{2}\left(
\begin{array}{c|c}
\begin{matrix}
f_1+g_{1}-\frac{f_{1LL}}{3}& 0&0 \\
0&f_1+\tfrac{2 f_{1LL}}{3}& 0\\
0&0&f_1-g_{1}-\tfrac{f_{1LL}}{3}
\end{matrix}
&
\begin{matrix}
0&\sqrt{2}\left(h_1-ih_{1LT}\right)&0 \\
 0& 0& \sqrt{2}\left(h_1+ih_{1LT}\right) \\
 0&0&0
 \end{matrix}
 \\ \hline
 \begin{matrix}
 0&0&0\\
\sqrt{2}\left(h_1+ih_{1LT}\right)& 0 &0\\
0& \sqrt{2}\left(h_1-ih_{1LT}\right)&0
 \end{matrix}
 &
\begin{matrix}
f_1-g_{1}-\tfrac{f_{1LL}}{3}& 0\\
 0& f_1+\tfrac{2f_{1LL}}{3}&0\\
 0&0&f_1+g_{1}-\tfrac{f_{1LL}}{3}
 \end{matrix}
\end{array}\right),
\end{equation}

\noindent where the elements of the \( 3 \times 3 \) block matrices refer to the \(\rho\)-helicity \( (\Lambda^\prime,\Lambda) \), with  $\Lambda',\Lambda=+1,0,-1$, whereas different blocks belong to different combinations of quark helicities \((\lambda^\prime,\lambda ) \), with $\lambda',\lambda=+,-$.
\\
For a spin $0$ target like the $\pi$ one has only one independent helicity amplitude, corresponding to the unpolarized PDF, i.e.
\begin{equation}\label{amplPDFs0}
f^{q/\pi}_1=A^{q/\pi}_{0+,0+}+A^{q/\pi}_{0-,0-}=2A^{q/\pi}_{0+,0+}.
\end{equation}

The convolution model for the helicity amplitudes reads

\begin{eqnarray}
A^{q/p}_{\Lambda^\prime\lambda^\prime,\Lambda\lambda}=A^{q/p,\, bare}_{\Lambda^\prime\lambda^\prime,\Lambda\lambda}+\delta A^{q/p}_{\Lambda^\prime\lambda^\prime,\Lambda\lambda},
\end{eqnarray}

where $A^{q/p,\, bare}$ is the contribution from the bare proton, described in terms of three-valence quarks, and  $\delta A^{q/p}$ is the contribution from the $BM$ fluctuation in the proton, which can be further decomposed as 

\begin{eqnarray}
\delta A^{q/p}_{\Lambda^\prime\lambda^\prime,\Lambda\lambda}=\sum_{B,M} \delta A^{q/BM}_{\Lambda^\prime\lambda^\prime,\Lambda\lambda}+\delta A^{q/MB}_{\Lambda^\prime\lambda^\prime,\Lambda\lambda},
\end{eqnarray}

with $\delta A^{q/BM}$ and $\delta A^{q/MB}$ corresponding to the active quark coming from the baryon  or meson, respectively.

Taking the appropriate combinations of the proton helicity amplitudes in the convolution model, giving the proton PDFs  according to \eqref{amplPDFs1/2}, and taking into account the relations \eqref{amplPDFs1} and \eqref{amplPDFs0}
between the meson helicity amplitudes and the PDFs, one can deduce the convolution model for the proton PDFs of Eqs.~\eqref{eq:splittingf1}, 
\eqref{eq:splittingg1} and \eqref{eq:splittingh1}.

%%%%%%%%%%%%%%%%%%%%%%%%%%%%%%%%%%%%%%%% REFERENCES %%%%%%%%%%%%%%%%%%%%%%%%%%%%%%%%%%%%%%%%%%%%%%%% 

\end{document}